\documentclass[fleqn,usenatbib]{mnras}

\usepackage{newtxtext,newtxmath}
\usepackage[T1]{fontenc}

\DeclareRobustCommand{\VAN}[3]{#2}
\let\VANthebibliography\thebibliography
\def\thebibliography{\DeclareRobustCommand{\VAN}[3]{##3}\VANthebibliography}

\usepackage{graphicx}	% Including figure files
\usepackage{amsmath}	% Advanced maths commands
\usepackage{mathtools}
\usepackage{hyperref}
\usepackage{xspace}	
\usepackage{euscript}

\usepackage[dvipsnames]{xcolor}	
\usepackage{pdflscape}
\usepackage{afterpage}
\usepackage{placeins}
\usepackage{bm}
\usepackage{ dsfont }
\usepackage{listings}
\usepackage{soul}
\usepackage{caption}
\usepackage{subcaption}
\usepackage{soul}
\usepackage{tabularx}

\usepackage{amsthm}

\usepackage{algorithm}
\usepackage[noend]{algpseudocode}

\definecolor{codegreen}{rgb}{0,0.6,0}
\definecolor{codegray}{rgb}{0.5,0.5,0.5}
\definecolor{codepurple}{rgb}{0.58,0,0.82}
\definecolor{backcolour}{rgb}{0.95,0.95,0.92}

\defcitealias{mwtrace1}{Paper I}

\newcommand{\gaia}{{\it Gaia}\xspace}

\definecolor{dkgreen}{rgb}{0,0.6,0}
\definecolor{gray}{rgb}{0.5,0.5,0.5}
\definecolor{mauve}{rgb}{0.58,0,0.82}

\definecolor{codegreen}{rgb}{0,0.6,0}
\definecolor{codegray}{rgb}{0.5,0.5,0.5}
\definecolor{codepurple}{rgb}{0.58,0,0.82}
\definecolor{backcolour}{rgb}{0.97,0.97,0.95}

\title[The Milky Way Photo-Astrometric Tracer Density]{The Photo-Astrometric Vertical Tracer Density of the Milky Way II: Results from \textit{Gaia}.}

\author[A. Everall, V. Belokurov, N. W. Evans, D. Boubert and R. Grand]{
Andrew Everall,$^{1}$\thanks{E-mail: asfe2@cam.ac.uk}
Vasily Belokurov$^{1}$,
N. Wyn Evans$^{1}$,
Douglas Boubert$^{2,3}$,
Robert J. J. Grand$^{4,5,6}$
\\
$^{1}$Institute of Astronomy, University of Cambridge, Madingley Road, Cambridge CB3 0HA, UK\\
$^{2}$Magdalen College, University of Oxford, High Street, Oxford OX1 4AU, UK\\
$^{3}$Rudolf Peierls Centre for Theoretical Physics, Clarendon Laboratory, Parks Road, Oxford OX1 3PU, UK\\
$^{4}$Max Planck Institute for Astrophysics, Karl-Schwarzschild-Str. 1, Postfach 1317, D-85741 Garching, Germany\\
$^5$Instituto de Astrof\'isica de Canarias, Calle Vía L\'actea s/n, E-38205 La Laguna, Tenerife, Spain\\
$^6$Departamento de Astrof\'isica, Universidad de La Laguna, Av. del Astrof\'isico Francisco S\'anchez s/n, E-38206, La Laguna, Tenerife, Spain\\
}

\date{Accepted XXX. Received YYY; in original form ZZZ}

\pubyear{2021}

\begin{document}
\label{firstpage}
\pagerange{\pageref{firstpage}--\pageref{lastpage}}
\maketitle

% Abstract of the paper
\begin{abstract}
%The tracer density of Milky Way stars is key to studying the formation history of our Galaxy and an important ingredient in dynamical models aiming to constrain the distribution of dark matter. It is also a primary goal of the \textit{Gaia} mission.
We use \textit{Gaia} photometry and astrometry to estimate the vertical spatial structure of the Milky Way at the Solar radius, formally accounting for sample incompleteness (the selection function) and parallax measurement uncertainty. Our results show impressive precision demonstrating the power of the \textit{Gaia} data. However, systematic errors dominate the parameter value uncertainties. We thoroughly test and quantify the impacts of all systematic uncertainties.
The vertical tracer density is modelled as a sum of two exponential profiles for the thin and thick discs, together with a spherically symmetric power-law  for the stellar halo.
We constrain the thin disc scale height as ${h_\mathrm{Tn}=260 \pm 3\, (\mathrm{stat}) \pm 26\,\mathrm{pc}\, (\mathrm{sys})}$ and thick disc ${h_\mathrm{Tk}=693 \pm 7 \,(\mathrm{stat}) \pm 121\,\mathrm{pc}\, (\mathrm{sys})}$. For the halo, we obtain a power law profile with $n_\mathrm{H}=3.543\pm0.023 \,(\mathrm{stat}) \pm0.259\, (\mathrm{sys})$. We infer a local stellar mass density for non-compact object stars of ${\rho_\mathrm{local}^* = 3.66\pm0.03\,(\mathrm{stat})\pm0.52 \times10^{-2}\,\mathrm{M}_\odot/\mathrm{pc}^3\,(\mathrm{sys})}$ and surface density of ${\Sigma_\mathrm{local}^* = 23.17\pm0.08\,(\mathrm{stat})\pm2.43\,\mathrm{M}_\odot/\mathrm{pc}^2\,(\mathrm{sys})}$. We find asymmetries above and below the disc with longer disc scale heights in the north but a flatter halo in the south at the $\lesssim 10$ per cent level.
%This work demonstrates the vast amount of information contained within the \textit{Gaia} catalogue however more general models are needed to fully exploit that data.
\end{abstract}

% Select between one and six entries from the list of approved keywords.
% Don't make up new ones.
%%%%%%%%%%%%%%%%%%%%%%%%%%%%%%%%%%%%%%%%%%%%%%%%%%

%%%%%%%%%%%%%%%%% BODY OF PAPER %%%%%%%%%%%%%%%%%%

\section{Introduction}

The 3D distribution of stars throughout the Milky Way is vital for understanding the formation history of our Galaxy. This `tracer density' is also a key ingredient in methods attempting to estimate the distribution of dark matter in the Milky Way, with important implications for both cosmological models and direct detection experiments \citep[e.g.,][]{Read2014}.

There is a rich history of research into the structure of the Milky Way from \citealt{Herschel1785}'s star-gage method to \citealt{Kapteyn1922}'s lens-shaped model to recent discoveries of asymmetries \citep[e.g.,][]{Widrow2012} and halo substructures~\citep[e.g.,][]{Belokurov2006}. However, across the vast majority of historical studies of Milky Way structure, the missing ingredient has been directly measured distances. As a result, we have typically been dominated by the statistical uncertainties regarding the distances to observed sources. \citealt{Kapteyn1922} even says
\textit{``I know of no more depressing thing in the whole domain of astronomy than to pass from the consideration of the accidental errors of our star places to that of their systematic errors.''}

The {\it Hipparcos} mission \citep{Perryman1997} dramatically improved the situation, providing milli-arcsecond precision parallax measurements and propelling the field forward \citep[e.g.][]{Creze1998}. However, the {\it Hipparcos} catalogue contains only $118\,000$ stars so the sample size is limited.

Countering the limited size of {\it Hipparcos}, large scale photometric surveys such as SDSS \citep{Gunn1998} and 2MASS \citep{Skrutskie2006} measured precise photometry for tens to hundreds of millions of sources. Using photometric colours and stellar evolution models, luminosities of stars can be estimated which are used to infer distances to stars. This has been used to infer the structure of the Milky Way \citep[e.g.][]{Robin2003, Bilir2006sdss, Bilir20062mass, Juric2008}.

This picture has entirely changed with the advent of \gaia. Its primary aim is to measure the spatial and velocity distribution of over a billion stars in the Galaxy~\citep{Prusti2016}. To do this, \gaia and DPAC\footnote{DPAC is the \gaia Data Processing and Analysis Consortium who we have to thank for producing the exquisite quality of data.} have now provided parallax measurements for 1,467,744,818 sources \citep{Brown2021} with precisions down to $10^{-2}$ mas \citep{Lindegren2021ast}. Given this quality of data, one would be forgiven for thinking a detailed 3D map of the stellar components of the Milky Way would be a straightforward task.

For reasons related to the completeness of \gaia astrometry and awkwardness of distance uncertainties, inferring the true spatial distribution of stars from \gaia is a difficult statistical problem. \citealt{Rix2021} provides a detailed introduction and discussion of the importance and challenges of estimating and applying selection functions in source density models. To avoid these issues, some studies using \gaia data to infer the structure of the Milky Way only work with subsets and infer distances from photometry \citep{Deason2019, Iorio2018}. Some elect to not use \gaia data at all \citep{Mateu2018, Fukushima2019, Dobbie2020}.

In \citealt[][henceforth \citetalias{mwtrace1}]{mwtrace1}, we introduce and validate a method to attack these problems head-on, accounting for the selection function of the \gaia astrometry sample and parallax uncertainties. In this work, we leverage \citetalias{mwtrace1}'s method to estimate the scale height of the Milky Way thin and thick discs, the radial profile of the halo and the local number density of stars for each component. 
%Kapteyn may have been depressed to learn that \gaia has taken us into an era of astrometry where systematic errors can very easily dominate statistical measurements. 
We model the impacts of multiple systematic uncertainties to estimate the effect these may have on our final results.

In Section~\ref{sec:data}, we introduce our high latitude sample, which is extracted from \gaia early data release 3 (EDR3). We describe the cuts used to remove a small number of contaminants. Our model for the vertical tracer populations is then fit to the \gaia data and we describe the results in Section~\ref{sec:results}. There are various simplifications and approximations used in the method and model which could, in principle, bias the parameter fits. These are discussed and tested in Section~\ref{sec:systematics}. We explain how our tests are used to quantify statistical and systematic uncertainties in Section~\ref{sec:stat_and_sys}. Finally, we discuss the results in comparison with the literature values in Section~\ref{sec:discussion}. 

\section{Data}
\label{sec:data}

\begin{figure*}
    \centering
    \includegraphics[width=\textwidth]{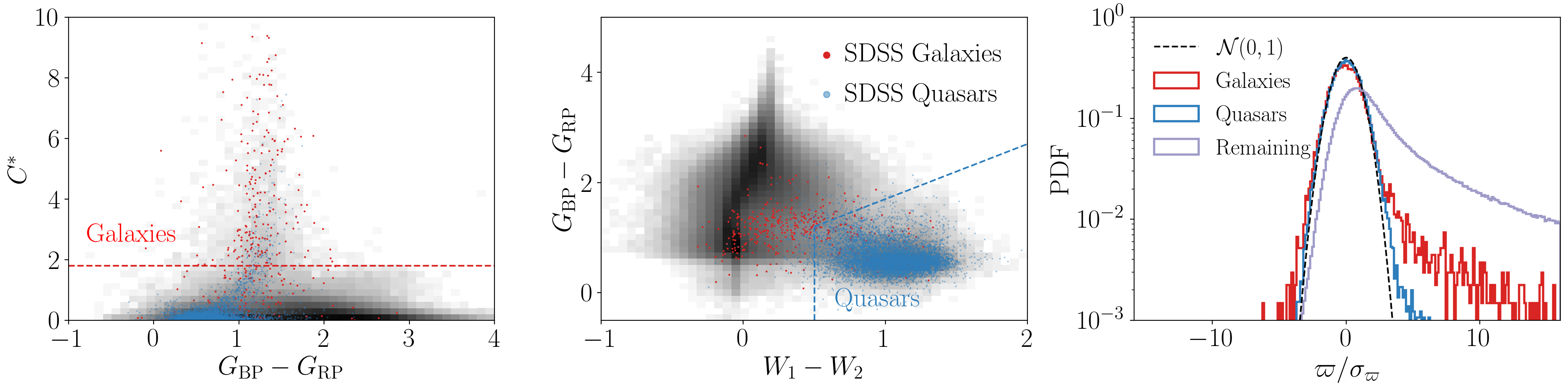}
    \caption{Cuts on \gaia and unWISE colour photometry are used to remove extragalactic sources from our sample. \textbf{\textit{Left}}: Galaxies are removed using a cut on \textsc{phot\_bp\_rp\_excess\_factor} after correcting for colour dependence, $C^*<1.8$. Red points show the SDSS spectroscopically classified galaxies which clearly extend to high excess flux levels. \textbf{\textit{Middle}}: Quasars are removed with colour-colour cuts on \gaia $G_\mathrm{BP}-G_\mathrm{RP}$ and unWISE $W_1-W_2$ shown by the blue dashed lines. The SDSS quasars (blue points) are clearly clustered in the region of colour-colour space beyond these cuts. \textbf{\textit{Right}}: The parallax SNR distribution of the galaxy and quasar samples are nearly Gaussian unit-variance distributed with a small enhancement at high $\varpi$ due to a small number of stars which are incorrectly removed from the sample. %\ev{for the right panel: perhaps log-scale the y axis and normalize the histograms to total number of stars/quasars/galaxies?} \andy{Good point with log scale, I've changed that. I want to keep it to PDF normalisation so that you get the comparison with a unit-variance normal distribution. I think this is less tidy if I scale it up.}
    }
    \label{fig:extragalactic}
\end{figure*}

Our initial sample of \gaia sources consists of all objects in EDR3 with $|b|>80^\circ$, published parallax with $\mathrm{RUWE}<1.4$ and published $G$-band apparent magnitude with $G>5$. Brighter sources saturate the \gaia CCDs which significantly affects the reliability of astrometric solutions. Our sample is extracted with the following query which returns 673\,926 sources in the Galactic north and 702\,599 in the south.
\begin{lstlisting}[language=SQL]
select ra, dec, parallax, parallax_error, phot_g_mean_mag from gaiaedr3.gaia_source 
where (b<-80 or b>80)
and parallax is not NULL
and phot_g_mean_mag>5
and ruwe<1.4
\end{lstlisting}

% Corrections
A recurring challenge with \gaia astrometry is the zero-point parallax offset, which leads to a small bias for any individual source but can significantly bias models fit to an entire population \citep[e.g. see][]{Everall2019}. We apply the zero-point correction recommended in \citealt{Lindegren2021plx} for sources with 5 and 6 parameter astrometric solutions. Many other groups have attempted to measure the zero point parallax offset from Cepheid variables \citep{Riess2021}, Red Clump stars \citep{Huang2021}, eclipsing binaries \citep{Stassun2021, Ren2021b} and quasars \citep{Groenewegen2021} (although the \citealp{Lindegren2021plx} model was constructed using quasars so it is unsurprising that these results match well). The conclusions are that for the majority of sources, the parallax offset is reduced to under $10\mu$as. \citealt{Zinn2021} and \citealt{Riess2021} find the parallaxes of sources brighter than $G=10.8$ are overestimated by $\sim15\mu$as after the correction, so we adjust the offset for the small portion of our sample with $G<10.8$. We test and discuss the effect of any residual offset in Section~\ref{sec:systematics}.

Parallax errors in \gaia are found to be typically underestimated when considering globular clusters \citep{Vasiliev2021} and wide binaries \citep{ElBadry2021}. We use the model from Equation (16) of \citealt{ElBadry2021} to revise the parallax errors of our \gaia sample, as this is appropriate for uncrowded fields which broadly applies to our sample. Close binary systems can bias the measured parallax for individual sources \citep{Belokurov2020, Penoyre2020} however this will predominantly affect nearby sources with specific orbital parameters and therefore not have a significant effect on our results.

The \gaia G-band apparent magnitude also has some small systematic bias for sources with 6-parameter astrometric solutions. We apply the apparent magnitude correction recommended in \citealt{Riello2021} to the sources where $G_\mathrm{BP}-G_\mathrm{RP}$ colour is available. One issue this raises is that the $G$-band apparent magnitude used for the data is subtly different from the measurements used to derive the \gaia selection function. However, the magnitude correction is at most $-0.025$ mag which is much smaller than our 0.2 mag resolution of the selection function. Therefore this inconsistency will have a negligible effect on the results.

% Extragalactic
As we are only using objects at high Galactic latitude, there is likely to be a sizable contamination from extragalactic sources (both quasars and distant galaxies). If left in the sample, these would bias the inferred distribution of stars towards larger distances.

Classifiers have been constructed to determine the probability of a source being extragalactic based on \gaia astrometry and photometry complemented with other surveys \citep{BailerJones2019, Shu2019}. The issue is that these classifications are not $100\%$ pure and will likely remove dim stars with low parallaxes which are misclassified as extragalactic. This is particularly clear in Fig.~10 of \citealt[][]{BailerJones2019} where the `quasar' population is dominated by the LMC, SMC and particular scans. The most prominent scans are the same as those found in Appendix B of \citealt[][]{CoGI} which were caused by missing calibration data in the \gaia photometric processing pipeline. To avoid introducing a bias to our data when removing extragalactic sources, we avoid selecting on apparent magnitude and astrometry.

Galaxies have an extended flux distribution on the sky. Due to the larger window size used to measure BP and RP on-board \gaia, galaxies will typically produce an excess flux in these bands over the $G$-band \citep[see Fig. 21][]{Riello2021}. The flux ratio between the combined BP and RP measurements and the $G$-band is published as $\textsc{phot\_bp\_rp\_excess\_factor}$ in the \gaia archive \citep{Evans2018}. The published excess flux has some residual colour-dependence which needs correcting. We use the formula provided in Section~6 of \citealt{Riello2021} to estimate the corrected flux excess $C^*$. Galaxies are selected as sources with $C^*>1.8$. The distribution of sources in excess flux vs $G_\mathrm{BP}-G_\mathrm{RP}$ is shown in the left panel Fig.~\ref{fig:extragalactic} with the red dashed line showing the Galaxy cut.

Quasars are well distinguished using the WISE photometry's $W_1 - W_2$ colour \citep[e.g.,][]{Shu2019}. We crossmatch our sample with the unWISE sample which has improved resolution over the original WISE catalogue \citep{Lang2014}. Taking the nearest object within 2~arcseconds correcting for proper motions with the \gaia epoch set to 2016 and unWISE to 2010 produces a successful match for 88\% of sources in our sample. Quasars are removed from our sample using the colour-colour cut
\begin{equation}
    W_1 - W_2 > 0.5 \quad \& \quad G_\mathrm{BP}-G_\mathrm{RP}<0.7(W_1-W_2)
\end{equation}
which is shown by the blue dashed line in the middle panel of Fig.~\ref{fig:extragalactic}.

These cuts select 2,933 galaxies and 50,726 quasars with 553 sources classified as both a galaxy and quasar. However, this does not tell us how successful our selection has been. For this, we crossmatch with spectroscopically classified sources in SDSS-IV \citep{Blanton2017}. We again use a proper motion corrected crossmatch for sources within 2~arcseconds with the SDSS epoch set at 2000. In this case, only 1.8\% of our sample receive SDSS spectra, the vast majority of which are in the northern field. The objects classified as galaxies and quasars by SDSS are shown as the red and blue points respectively in the left and middle panels of Fig.~\ref{fig:extragalactic}.

Of those with successful crossmatches, 8\,900 are classified as galaxies or quasars by SDSS whilst our cuts select 8\,275 sources, of which 8\,114 are classified as extragalactic by both. This implies that our selection criteria correctly classifies 91.2\% of extragalactic sources with only 1.7\% of Milky Way sources incorrectly classified as extragalactic. The remaining 8.8\% of missing sources account for $\sim0.3\%$ of our final sample, so we consider this completeness to be sufficient. 

Extragalactic sources are far too distant for \gaia parallax measurements therefore the measured parallax signal to noise will be distributed as $\varpi/\sigma_\varpi \sim \mathcal{N}(0,1)$. We show this distribution in the right hand panel of Fig.~\ref{fig:extragalactic} for galaxies (red), quasars (blue) and the remainder of the sample (purple). The extragalactic sources are close to normally distributed. The galaxy sample has a small amount of stellar contamination which marginally enhances the $+\varpi$ wing, but overall this shows that our classification has performed well.

\begin{figure}
    \centering
    \includegraphics[width=0.49\textwidth]{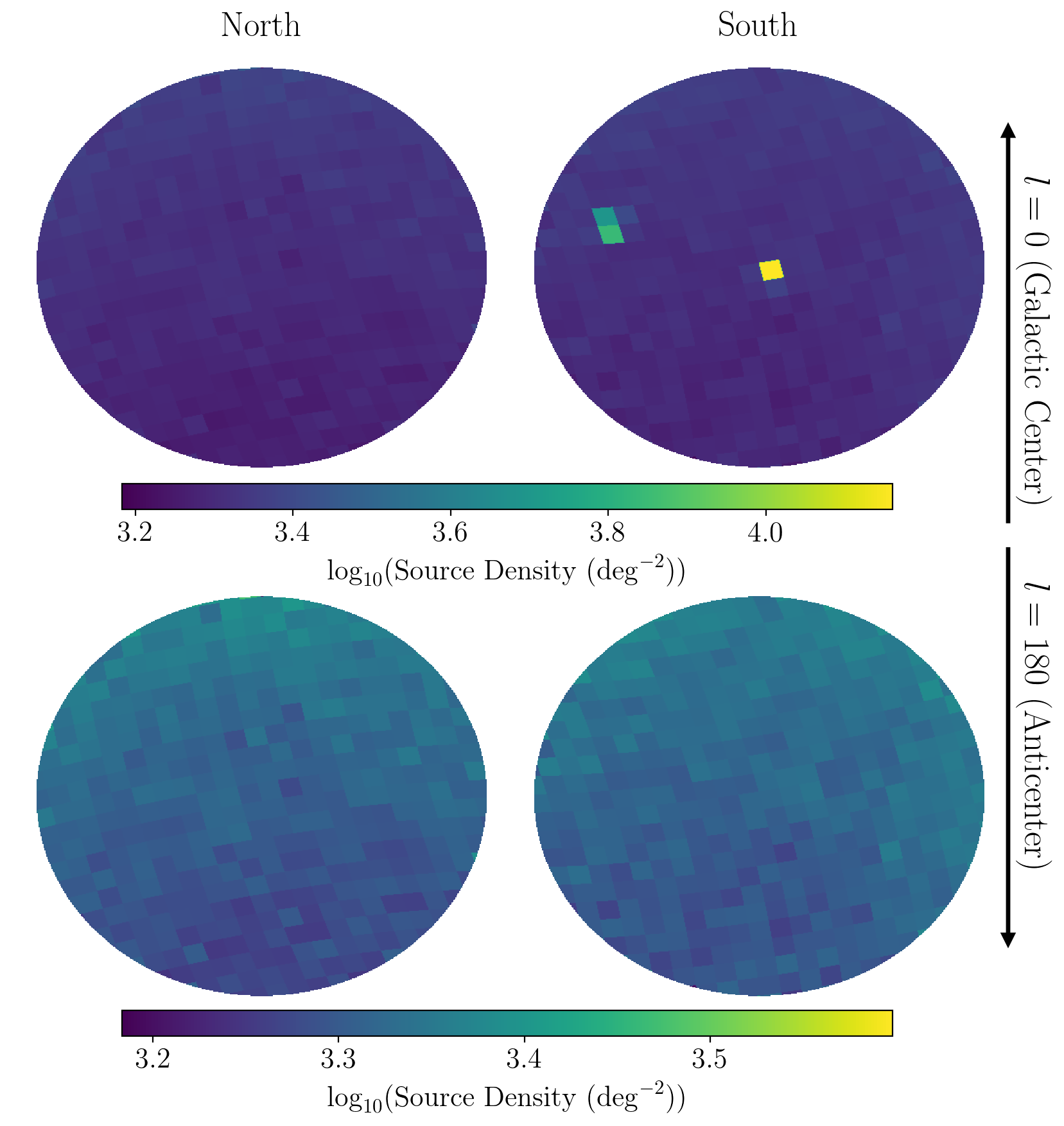}
    \caption{\textbf{\textit{Top}}: The number density of sources in HEALPix pixels across the north (left) and south (right) regions of the sky with $|b|>80^\circ$ is mostly uniform. The two clear exceptions are NGC 288 at the south Galactic pole and the Sculptor dwarf spheroidal at $(l,b)=(288^\circ,-83^\circ$) both appearing in the upper right panel. \textbf{\textit{Bottom}}: After masking these contributions we are left with the bottom panels which are almost uniform with a slight number density gradient from towards the Galactic center at the top to the anticenter at the bottom.}
    \label{fig:source_density}
\end{figure}
\begin{figure*}
    \centering
    \includegraphics[width=\textwidth]{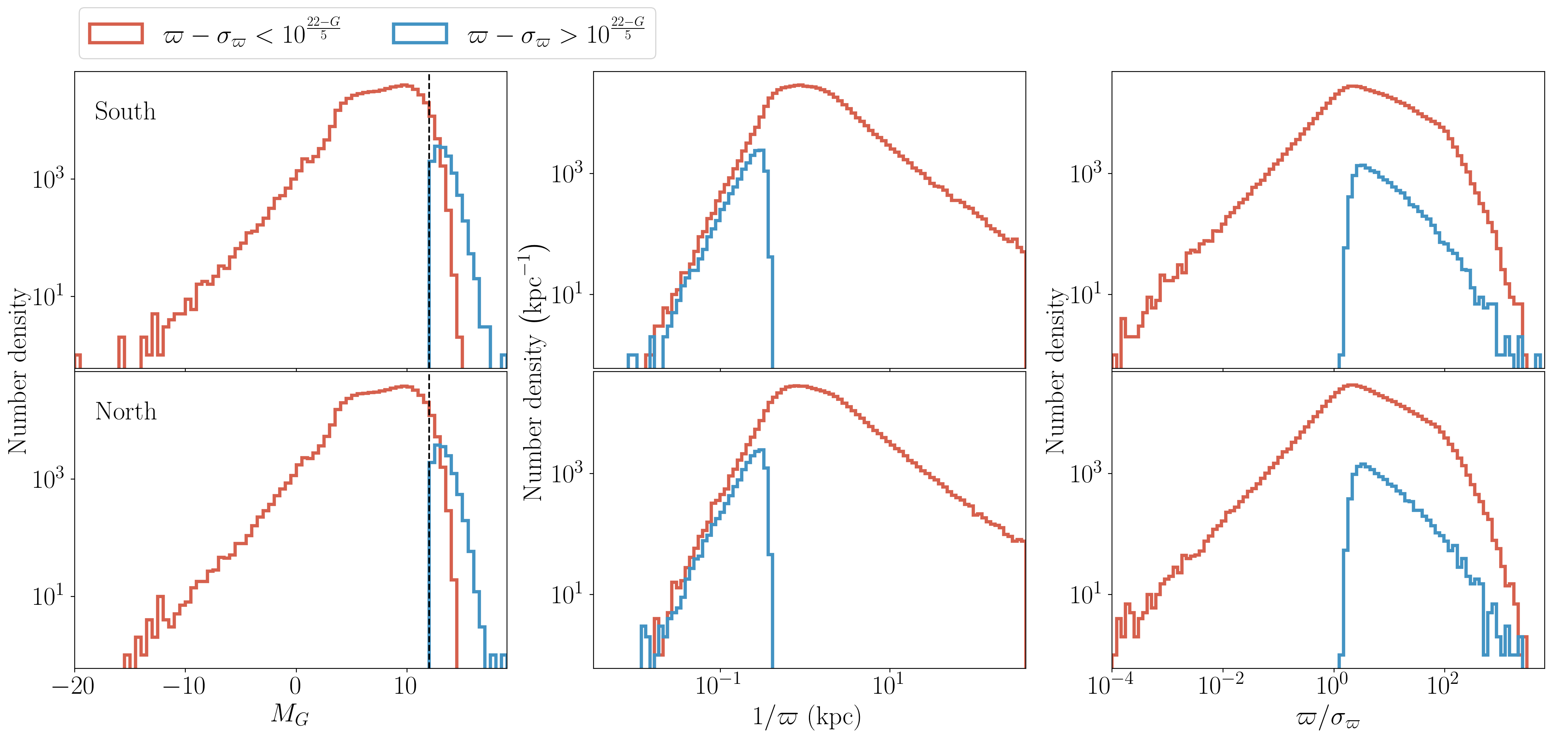}
    \caption{The effect of removing sources with $\varpi-\sigma_\varpi>10^{\frac{22-G}{5}}$ is shown as a function of absolute magnitude ($M_G=G+5\log_{10}(\varpi)-10$, left), inverse parallax (middle) and parallax signal-to-noise (right) for samples in the south (top) and north (bottom) regions. The cut conservatively removes sources which are likely to be intrinsically dimmer than the maximum absolute magnitude of the model ($M_G=12$). The removed sources (blue histograms) don't extend beyond $1/\varpi=400$ pc and all have parallax SNR greater than unity so we can be confident in their high absolute magnitudes. Some sources with $M_G>12$ will pass this very cautious cut but we expect these will be dominated by the number of bright sources with well measured parallax in the Solar neighbourhood. Our remaining samples after applying this cut are shown by the red histograms.}
    \label{fig:mg_cut}
\end{figure*}

% Substructure Masks
The number density of sources in pixels around the north and south Galactic poles is shown in the top panels of Fig.~\ref{fig:source_density}. For the most part, the distribution is reasonably smooth and noise dominated which is good when fitting a smooth model. However, the south field has two significant overdensities. The overdensity close to the south Galactic pole is the globular cluster NGC 288 which sits at a distance of approximately $9$ kpc from the Sun with a scale radius of $\sim 3$ arcminutes \citep{Vasiliev2021}. The other overdensity at slightly higher latitudes east of the Galactic Centre direction is the Sculptor dwarf spheroidal at $l=288^\circ,b=-83^\circ$ with a half-light radius of $\sim 11.3$ arcminutes~\citep{McConnachie2012}.

To prevent these objects from contaminating our smooth models, we mask the regions of the sky occupied by the structure out to four scale radii. We then renormalise the pixels by the fraction of the area which remains unmasked. This is the same treatment that we apply to pixels sitting on the edge of the $10^\circ$ radius fields. The resulting source density after masking NGC 288 and Sculptor is given by the bottom panels in Fig.~\ref{fig:source_density}, showing no further significant residual substructure. The gradient of the source density from the Galactic Centre (top of the figure) to the outer galaxy can now be seen. This shows the cylindrical radius dependence of the Milky Way distribution of stars which is not factored into our model, but we discuss its impact in Section~\ref{sec:systematics}.

The absolute magnitude model defined in Section~3.2 of \citetalias{mwtrace1} is limited by $M_G<12$ in order to avoid use of uncertain stellar evolutionary models. The \gaia sample may still contain sources dimmer than this limit, which are nonetheless near enough that \gaia is able to detect them. The issue is that we cannot directly measure absolute magnitude and parallax error is large enough for many sources that they will be scattered to that region of absolute magnitude space independent of their true brightness. Our compromise is to cut out sources which are likely to be fainter than $M_G=12$ by $1\sigma$ uncertainty in parallax. %\ev{I see now why the SF is a function of parallax (and plx error!) - but isn't the additional complexity in the normalization integral greater than the possible biases from including uncertain stellar evolution models for faint dwarfs?} \andy{Sorry, I think my description in paper 1 was a bit confusing. The selection function is not a function of parallax or parallax error. Yes this selection for $M_G>12$ adds some dependence but this only actually cuts out sources with high parallax SNR so shouldn't be a significant issue.}.  
In other words, removing all sources with greater than $84\%$ likelihood of $M_G>12$. This means only keeping sources with 
\begin{equation}
    \varpi - \sigma_\varpi < 10^{\frac{22-G}{5}}.
\end{equation}
The effect of this cut is shown in Fig.~\ref{fig:mg_cut}. The left panel shows the naive absolute magnitude distribution calculated with $s=1/\varpi$. The cut removes a large fraction of objects which fall outside the boundary. Importantly, from the middle and right panels, all of the sources removed from the sample are measured with $1/\varpi$ within $400$pc of the Sun with a parallax SNR greater than $1.4$. Any error in this cut will introduce a dependence of the selection function on measured parallax and parallax error. However, given the high parallax SNR of the removed sources, we expect that this dependence should be negligibly small. An added benefit of the cut we have placed here is that it will likely remove sources with poor astrometric solutions as classified by \citealt{Rybizki2021} and \citealt{Smart2021} which are typically fainter than $M_G=12$. This cut removes a further 13\,792 and 13\,731 sources from the north and south fields respectively.

After all of the cleaning, we are left with 633\,289 north and 640\,072 south sources in our sample. We emphasise that, through all of these cuts, we remove less than $11\%$ of the sample with published parallax, $G$ apparent magnitude and $\mathrm{RUWE}<1.4$. By comparison, a cut on $\varpi>0$ \textit{alone} (which is a serious crime, according to \citealt{Luri2018}) removes over $15\%$ and a signal-to-noise cut of $\varpi/\sigma_\varpi>4$ removes over $61\%$. We are modelling the vast majority of \gaia sources using the reliable astrometric and photometric data that is available.

\begin{figure*}
  \centering
  \begin{subfigure}[b]{\linewidth}
    \includegraphics[width=\linewidth]{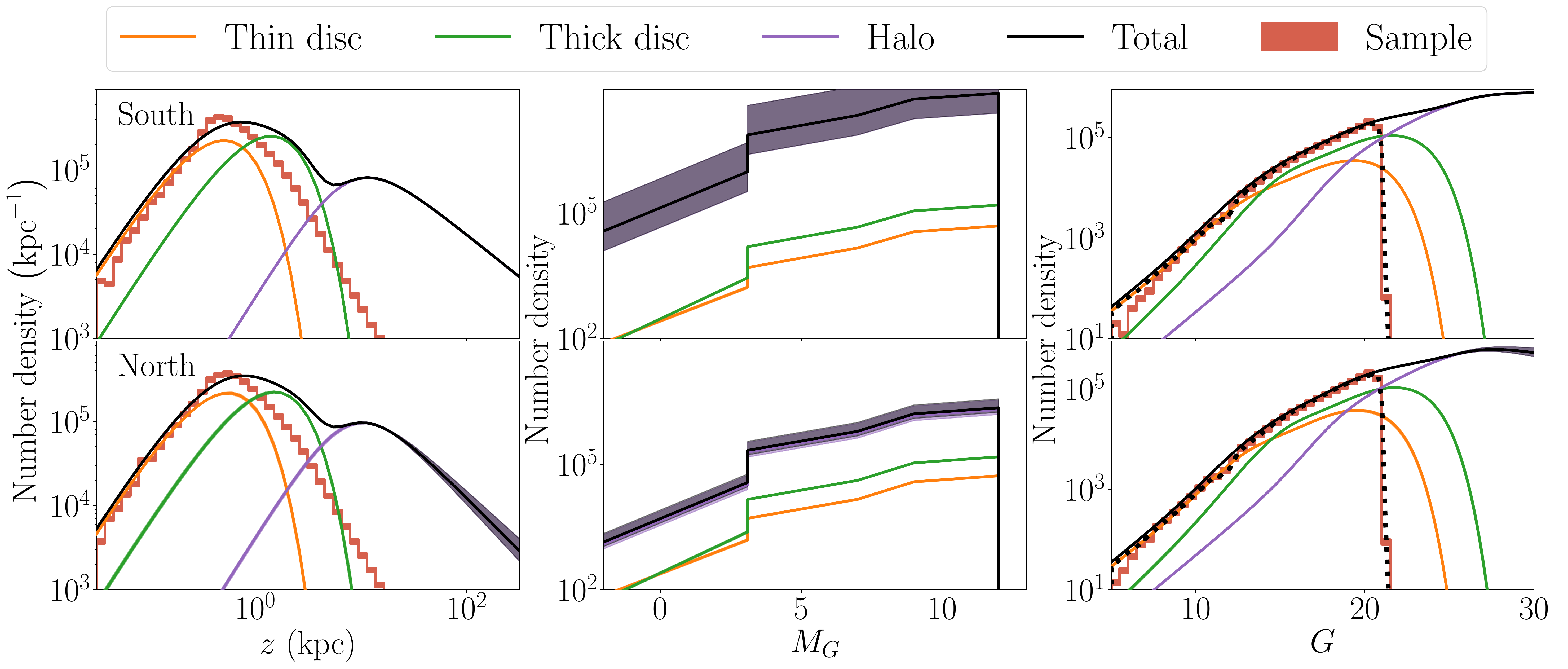}
    \caption{No Truncation}
    \label{fig:gaia_zMGa}
    \end{subfigure}
    
  \begin{subfigure}[b]{\linewidth}
    \includegraphics[width=\linewidth]{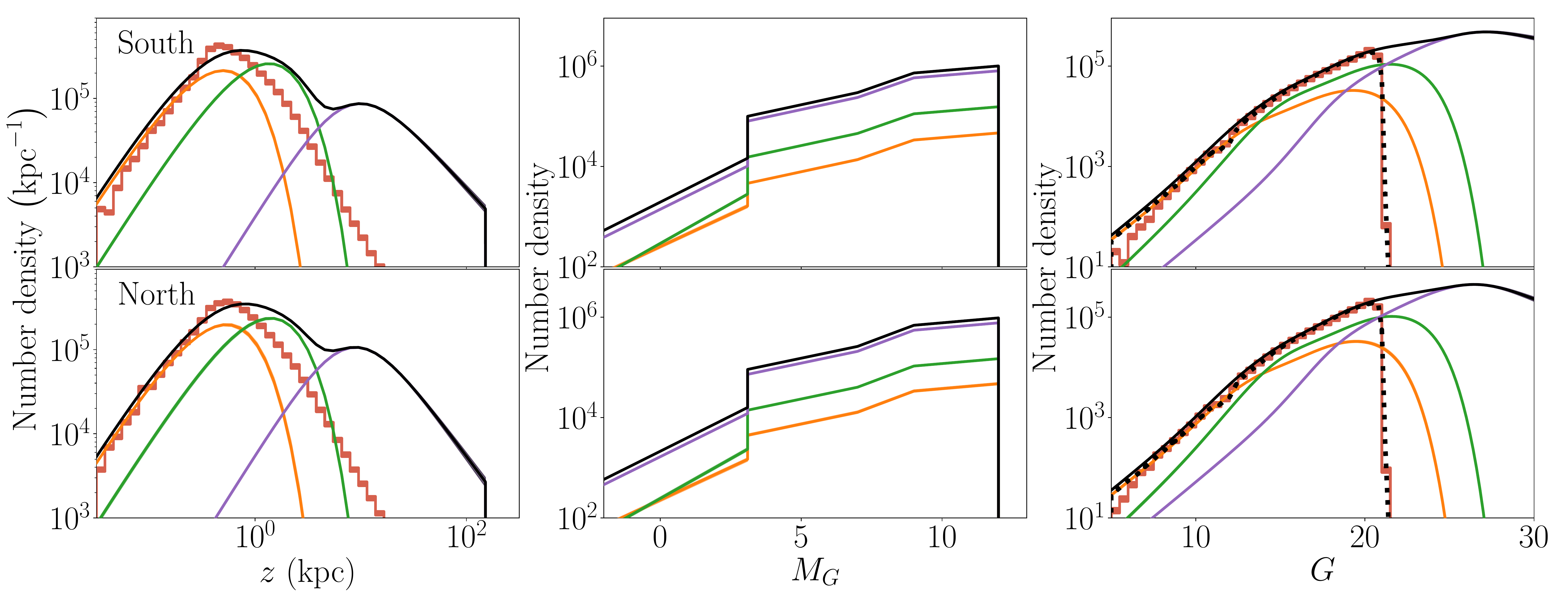}
    \caption{$s<160$ kpc}
    \label{fig:gaia_zMGb}
    \end{subfigure}
  \caption[]{Fitted models and sample number densities per unit $z$ (kpc, left), $M_G$ (middle) and $G$ (right) for the thin disc (orange) and thick disc (green) and halo (purple) and their sum total (black). Lines show the median model fits with shaded regions providing the $1^\mathrm{st}-99^\mathrm{th}$ percentile range of the posterior fits to the \gaia data. In most cases the posterior is so tightly constrained that the uncertainties cannot be picked out in these plots. \textbf{a}: For the infinite halo model there is qualitative agreement between south (top) and north (bottom) disc samples with a steeper northern halo profile. Due to the large total normalisation of the infinite halo within $b>80^\circ$, the halo dominates the absolute magnitude profile and it sits directly under the total profile. \textbf{b}: The model with halo truncated such that $s<160$ kpc also has similar north and south profiles with a marginally steeper south halo. For both models the red histograms in the left panels show the distribution $\sin(|b|)/\varpi$ which is significantly different to the fit model due to a combination of the selection function and parallax error which we have demonstrated need to be treated properly ($\sim14$ per cent of the sample has negative parallax and cannot even be plotted). Red histograms in the right column show the $G$ distribution of the data which agrees very well with the product of our model with the selection function (black dotted line). At the bright end the model slightly overestimates the data which is likely because our model does not truncate at the tip of the red giant branch (see Fig.~3 \citetalias{mwtrace1}).
  }
   \label{fig:gaia_zMG}
\end{figure*}

\begin{figure*}
  \centering
  \includegraphics[width=\textwidth]{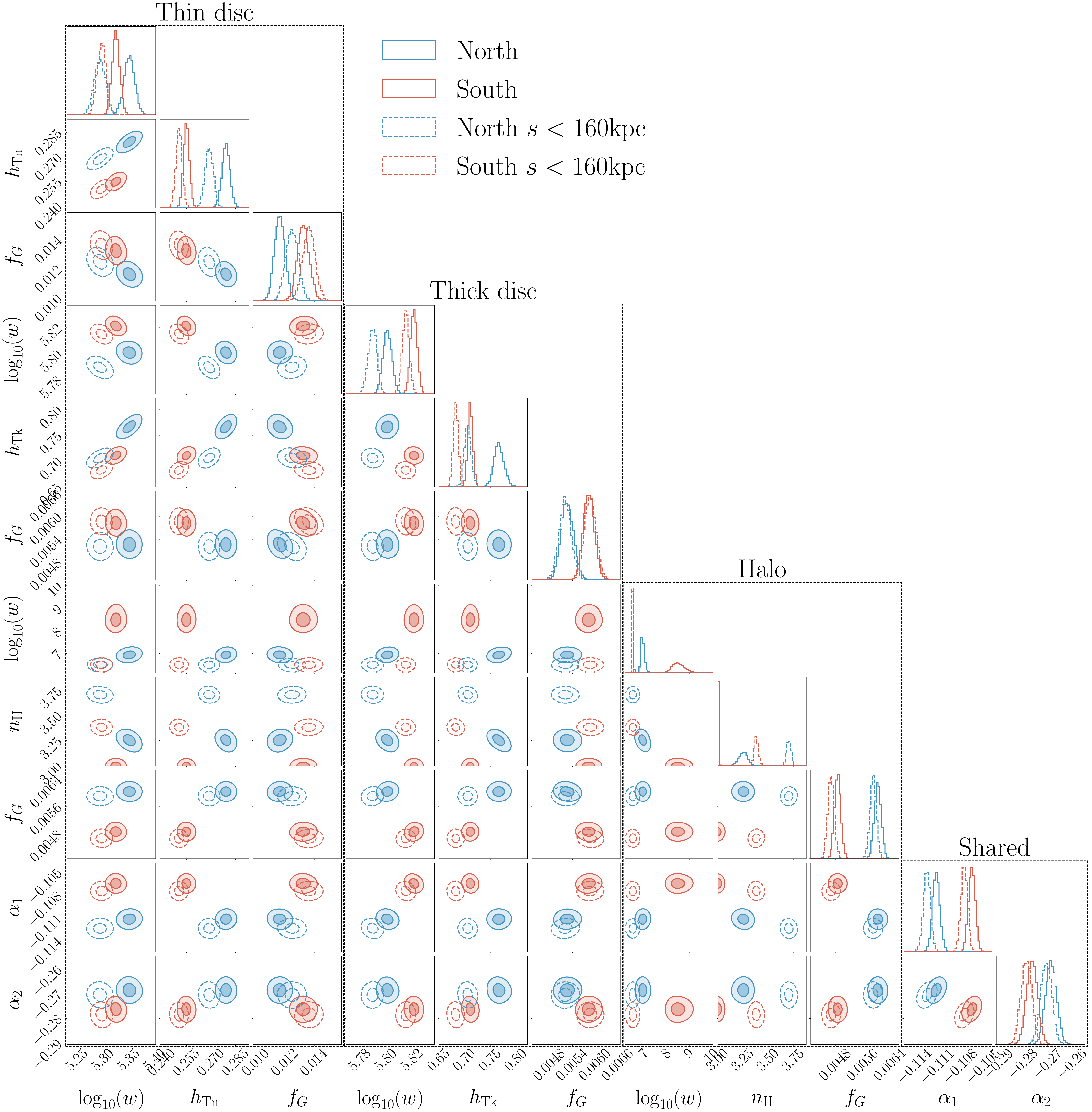}
  \caption[]{The posterior distributions for the north (red solid contours) and south (blue solid contours) sample fits show a small but significant disagreement across most parameters suggesting a weak asymmetry. Constraining the model to $s<160$ kpc (dashed contours) has a small impact on disc parameters however the halo model is much more significantly affected. Notably, the halo power-law index, which pushes close to the lower bound for an un-truncated model, is fit with a significantly steeper profile when the truncation is applied. The truncated model is better suited to the Milky Way for which the halo will not extend indefinitely.}
   \label{fig:gaia_corner}
\end{figure*}

\renewcommand{\arraystretch}{1.5}
%\begin{center}
\begin{table*}
\begin{tabular}{c c c c c c c c} 
 \hline
 Component & Parameter & North & South & North ($s<160$kpc) & South ($s<160$kpc) \\ [0.5ex] 
\hline\hline 
 Thin disc
 & $w$ 
 & ${2.24}_{-0.05}^{+0.05}\times 10^{5}$ 
 & ${2.11}_{-0.03}^{+0.04}\times 10^{5}$ 
 & ${1.97}_{-0.05}^{+0.05}\times 10^{5}$ 
 & ${1.98}_{-0.04}^{+0.04}\times 10^{5}$\\
 & $h_\mathrm{Tn}$ 
 & ${0.279}_{-0.002}^{+0.002}$ 
 & ${0.255}_{-0.002}^{+0.002}$ 
 & ${0.269}_{-0.002}^{+0.003}$ 
 & ${0.250}_{-0.002}^{+0.002}$\\
 & $f_G$ 
 & ${1.16}_{-0.04}^{+0.04}\times 10^{-2}$ 
 & ${1.32}_{-0.04}^{+0.04}\times 10^{-2}$ 
 & ${1.25}_{-0.04}^{+0.04}\times 10^{-2}$ 
 & ${1.36}_{-0.04}^{+0.04}\times 10^{-2}$\\
 & $M_\mathrm{TO}$ & 3.1 & & &\\
 & $\alpha_3$ & -0.6 & & & \\
 \hline\hline
 Thick disc
 & $w$ 
 & ${6.32}_{-0.05}^{+0.05}\times 10^{5}$ 
 & ${6.63}_{-0.04}^{+0.04}\times 10^{5}$ 
 & ${6.16}_{-0.05}^{+0.05}\times 10^{5}$ 
 & ${6.53}_{-0.04}^{+0.04}\times 10^{5}$\\
 & $h_\mathrm{Tk}$ 
 & ${0.766}_{-0.009}^{+0.010}$ 
 & ${0.711}_{-0.005}^{+0.005}$ 
 & ${0.706}_{-0.007}^{+0.007}$ 
 & ${0.683}_{-0.005}^{+0.005}$\\
 & $f_G$ 
 & ${5.26}_{-0.16}^{+0.17}\times 10^{-3}$ 
 & ${5.83}_{-0.14}^{+0.15}\times 10^{-3}$ 
 & ${5.21}_{-0.16}^{+0.16}\times 10^{-3}$ 
 & ${5.87}_{-0.15}^{+0.15}\times 10^{-3}$\\
 & $M_\mathrm{TO}$ & 3.1 & & &\\
 & $\alpha_3$ & -0.73 & & & \\
 \hline\hline
 Halo
 & $w$ 
 & ${8.64}_{-1.07}^{+1.69}\times 10^{6}$ 
 & ${3.26}_{-1.38}^{+3.11}\times 10^{8}$ 
 & ${3.19}_{-0.05}^{+0.05}\times 10^{6}$ 
 & ${3.39}_{-0.05}^{+0.06}\times 10^{6}$\\
 & $n_\mathrm{H}$ 
 & ${3.254}_{-0.046}^{+0.041}$ 
 & ${3.005}_{-0.003}^{+0.004}$ 
 & ${3.705}_{-0.023}^{+0.023}$ 
 & ${3.382}_{-0.020}^{+0.020}$\\
 & $f_G$ 
 & ${6.04}_{-0.10}^{+0.10}\times 10^{-3}$ 
 & ${4.86}_{-0.09}^{+0.09}\times 10^{-3}$ 
 & ${5.91}_{-0.09}^{+0.09}\times 10^{-3}$ 
 & ${4.66}_{-0.09}^{+0.09}\times 10^{-3}$\\
 & $M_\mathrm{TO}$ & 3.1 & & &\\
 & $\alpha_3$ & -0.64 & & & \\
 \hline\hline
 Shared
 & $\alpha_1$ 
 & ${-0.1111}_{-0.0005}^{+0.0005}$ 
 & ${-0.1065}_{-0.0004}^{+0.0004}$ 
 & ${-0.1123}_{-0.0005}^{+0.0005}$ 
 & ${-0.1075}_{-0.0004}^{+0.0004}$\\
 & $\alpha_2$ 
 & ${-0.2685}_{-0.0023}^{+0.0023}$ 
 & ${-0.2763}_{-0.0022}^{+0.0022}$ 
 & ${-0.2705}_{-0.0023}^{+0.0022}$ 
 & ${-0.2785}_{-0.0023}^{+0.0022}$\\\hline\hline 

\end{tabular}
\caption{The median and $16^\mathrm{th}-84^\mathrm{th}$ percentile ranges of the posterior parameter distributions from fits to the \gaia samples are shown for the north and south samples for an infinitely extending model and for a halo truncated at $s=160$ kpc. Across all parameters there is a significant asymmetry between the results of fitting to the north and south samples. The south disc profiles are steeper than the north with a smaller scale height, however, the southern halo is significantly shallower than the northern halo even pushing up against the prior boundary for the un-truncated model.}
\label{tab:gaiaresults}
\end{table*}
%\end{center}

\section{Method overview}
\label{sec:model_recap}

The method and model used are described in detail in Sections 2,3 and 4 of \citetalias{mwtrace1}. Here we will briefly recall the important details.

A three component model is used with a thin disc, thick disc and halo where each component is the product of a spatial and absolute magnitude distribution
\begin{equation}
f(l,b,\varpi,M_G) = \sum_{c=\{\mathrm{Tn},\mathrm{Tk},\mathrm{H}\}} w_c\,\nu_c(l,b,\varpi,\boldsymbol{\psi}_\nu) \, \phi_c(M_G,\boldsymbol{\psi}_\phi).
\end{equation}
$w_c$ is the total number of stars in the given component within $b>80^\circ$ or $b<-80^\circ$ for the north and south samples. As discussed in \citetalias{mwtrace1}, this is a significant assumption as the disc is known to have formed over an extended period of time \citep[e.g. ][]{Snaith2015} and the halo is made up of multiple stellar populations \citep[e.g. ][]{Helmi2018, Belokurov2018, Belokurov2020splash}. \citet{Bovy2012nothick} and \citet{Mackereth2017} have demonstrated that the thin and thick discs are not distinct in spatial structure and metallicity and both can be modelled by a single continuous distribution of profiles. We discuss such a model and the challenges it presents in Section~\ref{sec:general_models}, however, for this work we apply the simpler model composed of distinct thin and thick disc profiles.

The thin and thick discs are exponential profiles in $|z|$ with scale heights $h_\mathrm{Tn}$ and $h_\mathrm{Tk}$. The halo is a spherically symmetric power law profile with exponent $-n_\mathrm{H}$.

The absolute magnitude distributions are four-part exponential profiles, one for the giants with $M_G<M_\mathrm{TO}$ (the turn-off magnitude) and three for main sequence dwarfs. Exponents, $\alpha_1, \alpha_2$ of the lower and upper main sequence respectively are fit but constrained to take the same values for all components. The middle component of the main sequence is constrained by the continuity condition on $M_G$ and the IMF. The giant exponent, $\alpha_3 = -0.60, -0.77, -0.64$ is fixed for the thin disc, thick disc and halo respectively. $M_\mathrm{TO}=3.1$ is also fixed for all components. The free and fixed parameters are all listed in Table~\ref{tab:gaiaresults}. Priors on all parameters are listed in Table~1 of \citetalias{mwtrace1}.

The model is fit to the data with the likelihood function introduced in Section~2 of \citetalias{mwtrace1}. This is a Poisson likelihood function \citep[Appendix B, ][]{seestar} accounting for the selection function of the observatory and marginalising over parallax uncertainty.

The selection function for the \gaia DR2 source catalogue was developed in \citealt{CoGII} making use of the calibrated \gaia scanning law \citep{CoGI, CoGIII}. In \citealt{CoGV} this was updated with the EDR3 nominal scanning law and selection functions were evaluated for subsets of \gaia EDR3 using the method developed in \citealt{Boubert2021}. We use the product of the selection functions for the \gaia EDR3 source catalogue and subset with measured parallax and $\mathrm{RUWE}<1.4$. This accurately describes the incompleteness of the samples we are modelling.

To optimize the results, we first run MCMC on the priors using \textsc{emcee} \citep{ForemanMackey2013} with 44 walkers, 100 step burn-in and 100 steps of sampling. Ten samples are drawn from the prior for gradient descent with L-BFGS-B \citep{lbfgsb} implemented in \textsc{scipy}. The maximum likelihood estimate with the highest likelihood is used to start a second MCMC process with a small Gaussian ball around the parameter values. 44 walkers and 5000 steps are used with the final 2500 steps at 5 step intervals taken for the posteriors.

This method is used for all fits in Sections~\ref{sec:results} and \ref{sec:systematics}.

\section{Results}
\label{sec:results}

\renewcommand{\arraystretch}{1.5}
%\begin{center}
\begin{table*}
\begin{tabular}{c c c c c c c} 
 \hline\hline
 & $\mu^\mathrm{North}$& $\mu^\mathrm{South}$& $\mu$& $\sigma_\mathrm{sys}^{\mathrm{N}/\mathrm{S}}$& $\sigma_\mathrm{sys}^{\mathrm{Tests}}$& \\
 \hline\hline
 $\rho^*_\mathrm{local} \, \left(\mathrm{M}_\odot/\mathrm{pc}^3\right)$ 
 & ${3.28} \pm {0.03}\times 10^{-2}$ 
 & ${4.05} \pm {0.04}\times 10^{-2}$ 
 & ${3.66} \pm {0.03}\times 10^{-2}$ 
 & $\pm {0.39}\times 10^{-2}$ 
 & $\pm {0.34}\times 10^{-2}$ & ($z_\odot$)\\
 $\Sigma^*_\mathrm{local} \, \left(\mathrm{M}_\odot/\mathrm{pc}^2\right)$ 
 & ${21.58} \pm {0.08}$
 & ${24.77} \pm {0.10}$
 & ${23.17} \pm {0.08}$
 & $\pm {1.59}$
 & $\pm {1.84}$ & ($z_\odot$)\\
 $\log_{10}\left(M^*_\mathrm{Halo}\,\left(\mathrm{M}_\odot\right)\right)$ 
 & ${8.97} \pm {0.01}$ 
 & ${8.74} \pm {0.01}$ 
 & ${8.86} \pm {0.01}$ 
 & $\pm {0.12}$ 
 & $\pm {0.15}$ & ($R$)\\
 \hline
 $\rho_\mathrm{Tn} \, \left(\mathrm{M}_\odot/\mathrm{pc}^3\right)$ 
 & ${2.86} \pm {0.03}\times 10^{-2}$ 
 & ${3.56} \pm {0.03}\times 10^{-2}$ 
 & ${3.21} \pm {0.03}\times 10^{-2}$ 
 & $\pm {0.35}\times 10^{-2}$ 
 & $\pm {0.38}\times 10^{-2}$ & ($z_\odot$)\\
 $\rho_\mathrm{Tk} \, \left(\mathrm{M}_\odot/\mathrm{pc}^3\right)$ 
 & ${4.19} \pm {0.13}\times 10^{-3}$ 
 & ${4.92} \pm {0.11}\times 10^{-3}$ 
 & ${4.57} \pm {0.13}\times 10^{-3}$ 
 & $\pm {0.34}\times 10^{-3}$ 
 & $\pm {2.17}\times 10^{-3}$ & ($A_V$,$\sigma_G$)\\
 $\rho_\mathrm{H} \, \left(\mathrm{M}_\odot/\mathrm{pc}^3\right)$ 
 & ${2.13} \pm {0.03}\times 10^{-5}$ 
 & ${1.51} \pm {0.02}\times 10^{-5}$ 
 & ${1.82} \pm {0.03}\times 10^{-5}$ 
 & $\pm {0.31}\times 10^{-5}$ 
 & $\pm {0.22}\times 10^{-5}$ & ($R$)\\
 \hline
 $\rho_\mathrm{Tk}/\rho_\mathrm{Tn}$ 
 & ${0.147} \pm {0.005}$ 
 & ${0.138} \pm {0.003}$ 
 & ${0.141} \pm {0.005}$ 
 & $\pm {0.000}$ 
 & $\pm {0.075}$ & ($A_V$,$\sigma_G$)\\
 $\rho_\mathrm{H}/\rho_\mathrm{Tn}$ 
 & ${7.46} \pm {0.14}\times 10^{-4}$ 
 & ${4.25} \pm {0.07}\times 10^{-4}$ 
 & ${5.85} \pm {0.14}\times 10^{-4}$ 
 & $\pm {1.60}\times 10^{-4}$ 
 & $\pm {1.10}\times 10^{-4}$ & ($R$)\\
 \hline
 $\Sigma_\mathrm{Tn} \, \left(\mathrm{M}_\odot/\mathrm{pc}^2\right)$ 
 & ${15.36} \pm {0.15}$ 
 & ${17.82} \pm {0.13}$ 
 & ${16.59} \pm {0.15}$ 
 & $\pm {1.22}$ 
 & $\pm {2.22}$ & ($z_\odot$)\\
 $\Sigma_\mathrm{Tk} \, \left(\mathrm{M}_\odot/\mathrm{pc}^2\right)$ 
 & ${5.92} \pm {0.13}$ 
 & ${6.72} \pm {0.11}$ 
 & ${6.33} \pm {0.13}$ 
 & $\pm {0.38}$ 
 & $\pm {1.67}$ & ($\sigma_G$,$A_V$)\\
 $\Sigma_\mathrm{H} \, \left(\mathrm{M}_\odot/\mathrm{pc}^2\right)$ 
 & ${0.294} \pm {0.003}$ 
 & ${0.225} \pm {0.002}$ 
 & ${0.259} \pm {0.003}$ 
 & $\pm {0.034}$ 
 & $\pm {0.019}$ & ($R$)\\
 \hline
 $h_\mathrm{Tn} \, (\mathrm{kpc})$ 
 & ${0.269} \pm {0.003}$ 
 & ${0.250} \pm {0.002}$ 
 & ${0.260} \pm {0.003}$ 
 & $\pm {0.009}$ 
 & $\pm {0.024}$ & ($\sigma_G$,$A_V$)\\
 $h_\mathrm{Tk} \, (\mathrm{kpc})$ 
 & ${0.706} \pm {0.007}$ 
 & ${0.682} \pm {0.005}$ 
 & ${0.693} \pm {0.007}$ 
 & $\pm {0.010}$ 
 & $\pm {0.121}$ & ($R$)\\
 $n_\mathrm{H}$ 
 & ${3.705} \pm {0.023}$ 
 & ${3.382} \pm {0.020}$ 
 & ${3.543} \pm {0.023}$ 
 & $\pm {0.160}$ 
 & $\pm {0.204}$ & ($R$)\\
 \hline
 $f^\mathrm{G}_\mathrm{Tn}$ 
 & ${1.25} \pm {0.04}\times 10^{-2}$ 
 & ${1.36} \pm {0.04}\times 10^{-2}$ 
 & ${1.31} \pm {0.04}\times 10^{-2}$ 
 & $\pm {0.04}\times 10^{-2}$ 
 & $\pm {0.84}\times 10^{-2}$ & ($M_\mathrm{TO}^\mathrm{Tn}$,$\sigma_G$)\\
 $f^\mathrm{G}_\mathrm{Tk}$ 
 & ${5.21} \pm {0.16}\times 10^{-3}$ 
 & ${5.87} \pm {0.15}\times 10^{-3}$ 
 & ${5.55} \pm {0.16}\times 10^{-3}$ 
 & $\pm {0.29}\times 10^{-3}$ 
 & $\pm {1.05}\times 10^{-3}$ & ($\Delta\varpi$,$A_V$)\\
 $f^\mathrm{G}_\mathrm{H}$ 
 & ${5.91} \pm {0.10}\times 10^{-3}$ 
 & ${4.67} \pm {0.09}\times 10^{-3}$ 
 & ${5.28} \pm {0.10}\times 10^{-3}$ 
 & $\pm {0.61}\times 10^{-3}$ 
 & $\pm {0.64}\times 10^{-3}$ & ($\Delta\varpi$)\\
 \hline
 $\alpha_1$ 
 & ${-0.1123} \pm {0.0005}$ 
 & ${-0.1075} \pm {0.0004}$ 
 & ${-0.1099} \pm {0.0005}$ 
 & $\pm {0.0024}$ 
 & $\pm {0.0050}$ & ($\Delta\varpi$)\\
 $\alpha2$ 
 & ${-0.2705} \pm {0.0023}$ 
 & ${-0.2785} \pm {0.0022}$ 
 & ${-0.2745} \pm {0.0023}$ 
 & $\pm {0.0034}$ 
 & $\pm {0.0244}$ & ($\Delta\varpi$)\\\hline\hline
\end{tabular}
\caption{Transformed results from the model fits to \gaia data are given for the north, south and combined samples along with one standard deviation uncertainties. While the statistical uncertainties for each sample are incredibly tight for most parameters, the systematic uncertainties due to north-south asymmetry and model oversimplifications are much larger. It is important to consider these additional systematics when using our results. The systematic uncertainties should be added in quadrature to the statistical uncertainty.}
\label{tab:transformed_results}
\end{table*}
%\end{center}

The model is independently fit to the northern and southern \gaia samples. This halves the sample size in either fit but means we can draw a comparison between the Milky Way structure above and below the disc. The method is described in detail in Section 2 of \citetalias{mwtrace1}.

The resultant model is shown in Fig.~\ref{fig:gaia_zMGa}. Solid lines and shaded regions show the median and $1^\mathrm{st}-99^\mathrm{th}$ percentile ranges for the fits to the individual components and sum total. We evaluate this by drawing 1000 samples from the MCMC posterior, evaluating the model and taking the percentiles as a function of $z$, $M_G$ and $G$. Each of the three Galaxy components are well constrained with the thin disc dominating the model for $z<0.5$ kpc, the thick disc being the main contribution for $0.5<z<5$ kpc and the halo taking over at large distances. The thin and thick disc profiles are qualitatively very similar between the north and south samples however the halo profile in the south fit declines much more slowly with distance.

In the left panels of Fig.~\ref{fig:gaia_zMGa} red histograms show the number density of stars as a function of $z=\sin(b)/\varpi$, which provides a biased estimate of height above the Milky Way disc. The distribution is significantly lower than our model at large scale heights both due to the selection function and because parallax uncertainty scatters measurements to either larger positive or negative observed parallax. Faint sources in \gaia have typical parallax uncertainties $\sigma_\varpi>0.1$ mas and so measuring $0<\varpi<0.1$ mas (which corresponds to $z>10$ kpc) is unlikely due purely to measurement noise. Some of these sources are scattered up in parallax and down in distance generating the excess of sources with measured $\varpi \sim 0.4$ kpc. This can also be seen in Figure 5 of \citetalias{mwtrace1} where the imposed \gaia-like selection function and parallax uncertainties have the same effect on the naive distribution of $z=\sin(b)/\varpi$. The point we are making here is that one must account for both parallax uncertainty and the \gaia selection function to obtain an unbiased model of the Milky Way distribution of stars.

%An interesting aspect of the fit is that the total weight is dominated by the halo. This is because, both in the north and south fits, the power law index of the halo has been pushed towards the lower end of possible values as shown in Fig.~\ref{fig:gaia_corner}. 
Unlike several previous works such as \citealt{Juric2008} and  \citealt{Mateu2018}, our model extends to infinity so we require $n_\mathrm{H}>3$ to keep the model normalisable. %\ev{as I commented in paper 1, this is probably redundant: all you need is the product of density times selection function to be normalizable.} \andy{Responded to this in Paper I. This normalisation is required for the given parameterisation.} 
However, this is unphysical and other studies have shown that the halo drops off much steeper beyond $r\gtrsim 50$ kpc \citep{Deason2014} or $r\sim 160$ kpc \citep{Fukushima2019}. Our model is dominated by information from the inner, shallower component of their profiles. This leads to an overestimate of the overall halo normalisation, which consequently are untrustworthy. 

To obtain a more realistic halo normalisation, we rerun the fits truncating the parallax integral and halo normalisation with $s<160$ kpc (i.e. $1/s>6.25\,\mu$as) and changing the halo exponent prior to $n_\mathrm{H}\sim\mathrm{U}[2,7.3]$. The spatial and absolute magnitude profiles are shown in Fig.~\ref{fig:gaia_zMGb}. In this case, the north and south halo profiles are both significantly steeper. 

The right hand panels of Figs.~\ref{fig:gaia_zMGa} and~\ref{fig:gaia_zMGb} show the apparent magnitude distribution marginalised over position on the sky and distance. We weight the total distribution by the selection function which produces the black dotted line. This sits directly on the red histograms which give the apparent magnitude distribution of the \gaia data. The model slightly overestimates the apparent magnitude distribution at the bright end ($G\lesssim 7$) which we expect is due to the truncation of the absolute magnitude distribution at the tip of the red giant branch which can be seen at $M_G\sim -3$ in Fig.~5 of \citetalias{mwtrace1} but which we do not account for in our model.

The posteriors on each parameter are shown in Fig.~\ref{fig:gaia_corner} for the north and south samples (blue and red respectively) with distance truncated fits shown with dashed contours. Across all parameters there are systematic differences between the results from the north and south samples. For the thin and thick disc parameters these differences are small. However, in the case of the halo, the effect is far more substantial. Transitioning from an infinite to a truncated halo also significantly modifies the halo parameters with small knock-on effects to the disc. Given previous work \citep{Deason2014, Fukushima2019}, we consider the truncated model to be the more appropriate and will use those fits for our final results.

%The parameter fits in the north and south samples, shown by the blue and red posterior distributions in Fig.~\ref{fig:gaia_corner} show strong consistency in many cases. Two exceptions are the scale heights of the thin and thick discs where the south sample returns a slightly but significantly higher value. There are two potential causes of this. Firstly, our model has assumed that $z_\odot=0$. In reality the sun is slightly offset from the plane to the northern Galactic side \citep{Binney1997}. This creates a small perspective effect which can cause sources in the south to appear to be distributed at greater distances than they are and in the north to be closer. We test the affect of shifting $z_\odot$ in more detail in Section~\ref{sec:systematics}. 

%Other potential causes are more physical. We already mentioned the two over-densities in the southern field, NGC 288 and SclD which are at large distances and can therefore slightly bias the overall fits. There is also known asymmetry between the northern and southern MW disc which could generate a similar effect. .

The posterior median, $16^\mathrm{th}$ and $84^\mathrm{th}$ percentiles for all components and parameters in each of the runs are given in Table~\ref{tab:gaiaresults}. 

\subsection{Stellar mass density}
\label{sec:results_mass}

Our parameterisation, in particular the component normalisation ($w_c$), is specific to this sample as it is the total number of source with $M_G<12$ within the cone $|b|>80^\circ$. The local stellar mass density ($\rho^*_\mathrm{local}$), local surface density ($\Sigma^*_\mathrm{local}$) and halo total stellar mass ($M_\mathrm{Halo}^*$) are more generally interesting to the Galactic dynamics community and can be estimated from our results as we will explain here.

The number density of sources in the Solar neighbourhood with $M_G<12$ is given by $w_c\cdot\nu_c(s=0)$ where subscript $c$ refers to each of the three Milky Way components. We can inflate this to include main sequence sources with $M_G>12$ using the isochrones from Section~3 of \citetalias{mwtrace1} and IMF. The isochrones translate $M_G=12$ to a minimum initial mass of sources in our sample for each component, giving $\mathcal{M}_\mathrm{ini,min}=0.177, 0.147, 0.115\,\mathrm{M}_\odot$ for the thin, thick disc and halo respectively. The maximum initial mass of stars before they reach the post-AGB evolution phase -- eventually leading to a compact object remnant and thus disappearing from our sample -- is $\mathcal{M}_\mathrm{ini,max}=1.083, 0.980, 0.801\,\mathrm{M}_\odot$. To get the total pre-compact object local number density of sources, we inflate our local number density by a factor
\begin{equation}
    X_c = \frac{\int_0^{\mathcal{M}_\mathrm{ini,max}} \xi(\mathcal{M}_\mathrm{ini}) \mathrm{d}\mathcal{M}_\mathrm{ini} }{\int_{\mathcal{M}_\mathrm{ini,min}}^{\mathcal{M}_\mathrm{ini,max}}\xi(\mathcal{M}_\mathrm{ini}) \mathrm{d}\mathcal{M}_\mathrm{ini}}
\end{equation}
where $\xi(\mathcal{M}_\mathrm{ini})$ is the IMF (we use \citealt{Kroupa2001}). This gives ${X_c=3.167,2.785,2.398}$ for the three components.

To estimate the local stellar mass density, we need the mean mass of sources in the population. We can use the IMF again for this however we need to account for stellar mass loss. We use the three component isochrones to transform from initial mass to current stellar mass, $\mathcal{M}(\mathcal{M}_\mathrm{ini})$. The stellar evolution models do not extend all the way to zero mass so we assume any stars with $\mathcal{M}_\mathrm{ini}<0.1$ experience negligible mass loss in their lifetimes such that ${\mathcal{M}(\mathcal{M}_\mathrm{ini})=\mathcal{M}_\mathrm{ini}}$.%Firstly we assume that stars with $\mathcal{M}_\mathrm{ini}<\mathcal{M}_\mathrm{ini,min}$ experience negligible mass loss such that $\mathcal{M}\approx \mathcal{M}_\mathrm{ini}$. For sources with  $\mathcal{M}_\mathrm{ini,min}<\mathcal{M}_\mathrm{ini}<\mathcal{M}_\mathrm{ini,max}$ we sample from the IMF and estimate the mean stellar mass, $\langle \widetilde{\mathcal{M}}_c\rangle$, of the sources in our sample  with $M_G<12$ using the isochrones. 
The mean mass of all non-compact object stars is
\begin{align}
    \langle \mathcal{M}\rangle_c = \frac{\int_0^{\mathcal{M}_\mathrm{ini,max}} \,\mathcal{M}(\mathcal{M}_\mathrm{ini})\,\xi(\mathcal{M}_\mathrm{ini})\, \mathrm{d}\mathcal{M}_\mathrm{ini}} {\int_0^{\mathcal{M}_\mathrm{ini,max}}\xi(\mathcal{M}_\mathrm{ini})\, \mathrm{d}\mathcal{M}_\mathrm{ini}}\nonumber
\end{align}
% \begin{align}
%     \langle &\mathcal{M}_c\rangle = \\ &\frac{\int_0^{\mathcal{M}_\mathrm{ini,max}} \,\mathcal{M}_\mathrm{ini}\,\xi(\mathcal{M}_\mathrm{ini})\, \mathrm{d}\mathcal{M}_\mathrm{ini} \,+\, \int_{\mathcal{M}_\mathrm{ini,min}}^{\mathcal{M}_\mathrm{ini,max}} \,\xi(\mathcal{M}_\mathrm{ini}) \,\mathrm{d}\mathcal{M}_\mathrm{ini}} {\int_0^{\mathcal{M}_\mathrm{ini,max}}\xi(\mathcal{M}_\mathrm{ini})\, \mathrm{d}\mathcal{M}_\mathrm{ini}}\nonumber
% \end{align}
which gives $\langle \mathcal{M}_c\rangle =0.174,0.168,0.155$ for the three components. Finally, the local mass density of non-compact object stars is
\begin{align}
    \rho^*_{\mathrm{local},c} = w_c \, \nu_c(s=0) \cdot X_c \cdot \langle \mathcal{M}\rangle_c.
\end{align}

A critical assumption we have made is that any stars born with an initial mass larger than $\mathcal{M}_\mathrm{ini,max}$ will not appear in our sample. In reality, the White Dwarf sequence extends up to $M_G\sim8$ \citep[see ][]{Rix2021} and so there may be many White Dwarfs in our sample. However, these will be dominated by the main sequence dwarfs of the same absolute magnitude and will only provide a severely sub-dominant contribution to the number density \citep[see Fig. 2][]{Smart2021}.

We estimate the surface densities by integrating our components with respect to $z$ with the mathematical working given in Appendix~\ref{app:surfacedensity}. Since our power-law halo has $n>3$, the total halo stellar mass is not well normalised at $r=0$. We estimate the total halo mass, $M^*_\mathrm{Halo}$, by integrating our halo profile for $r>1$ kpc and taking $n_H=2$ (uniform density) inside. As a result, the halo mass is largely dominated by stars inside the Solar radius and is an extrapolation of the local halo stellar mass density so this should be taken with caution.

The means and standard deviations of these parameters using the $s<160$ kpc fits are $\mu^\mathrm{North}$ and $\mu^\mathrm{South}$ in Table~\ref{tab:transformed_results}.

\section{Systematic Errors}
\label{sec:systematics}

\begin{figure*}
\begin{subfigure}[b]{.47\linewidth}
\includegraphics[width=\linewidth]{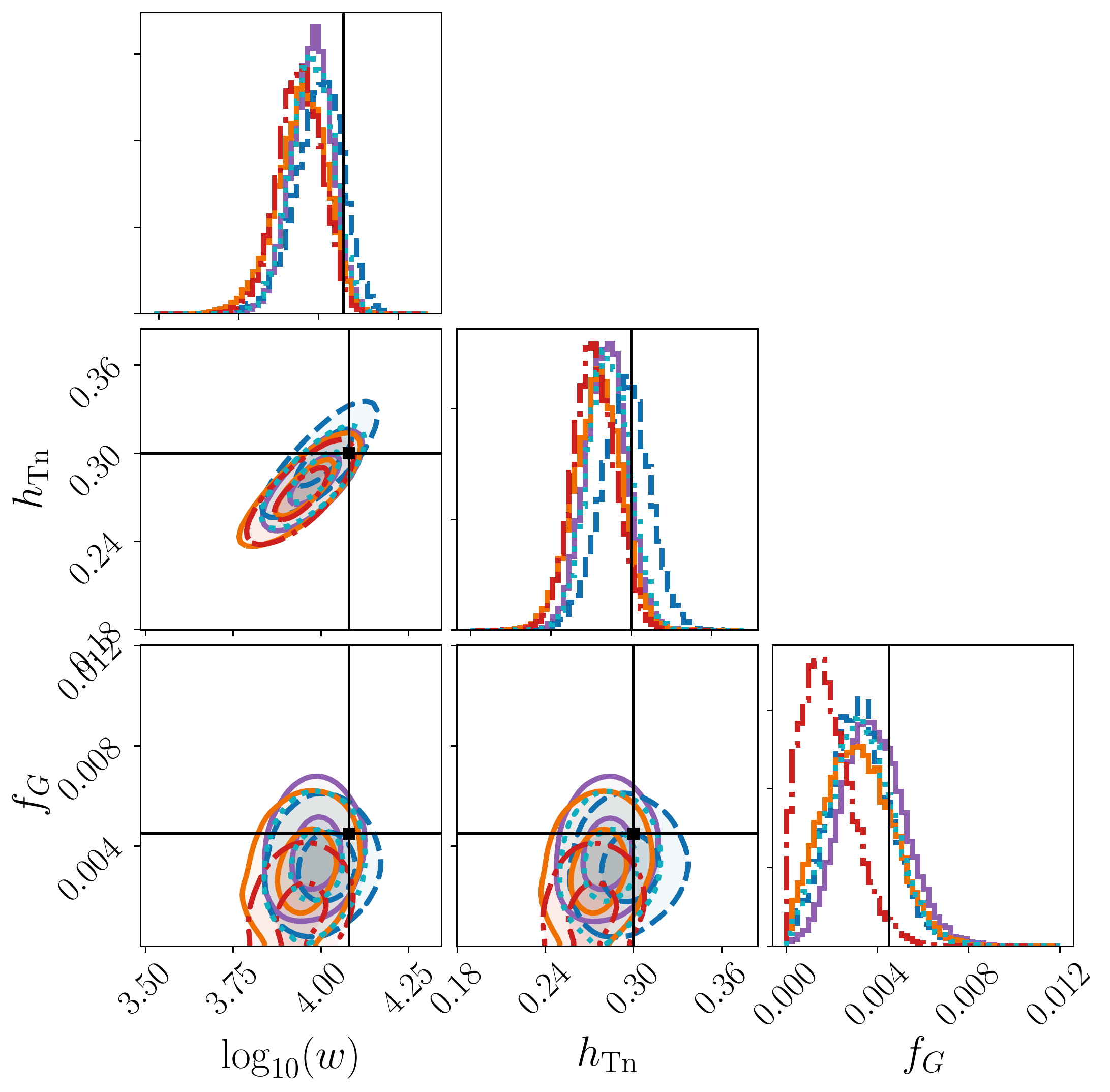}
\caption{Thin disc}
\label{fig:sys_thin_disk}
\end{subfigure}
\begin{subfigure}[b]{.47\linewidth}
\includegraphics[width=\linewidth]{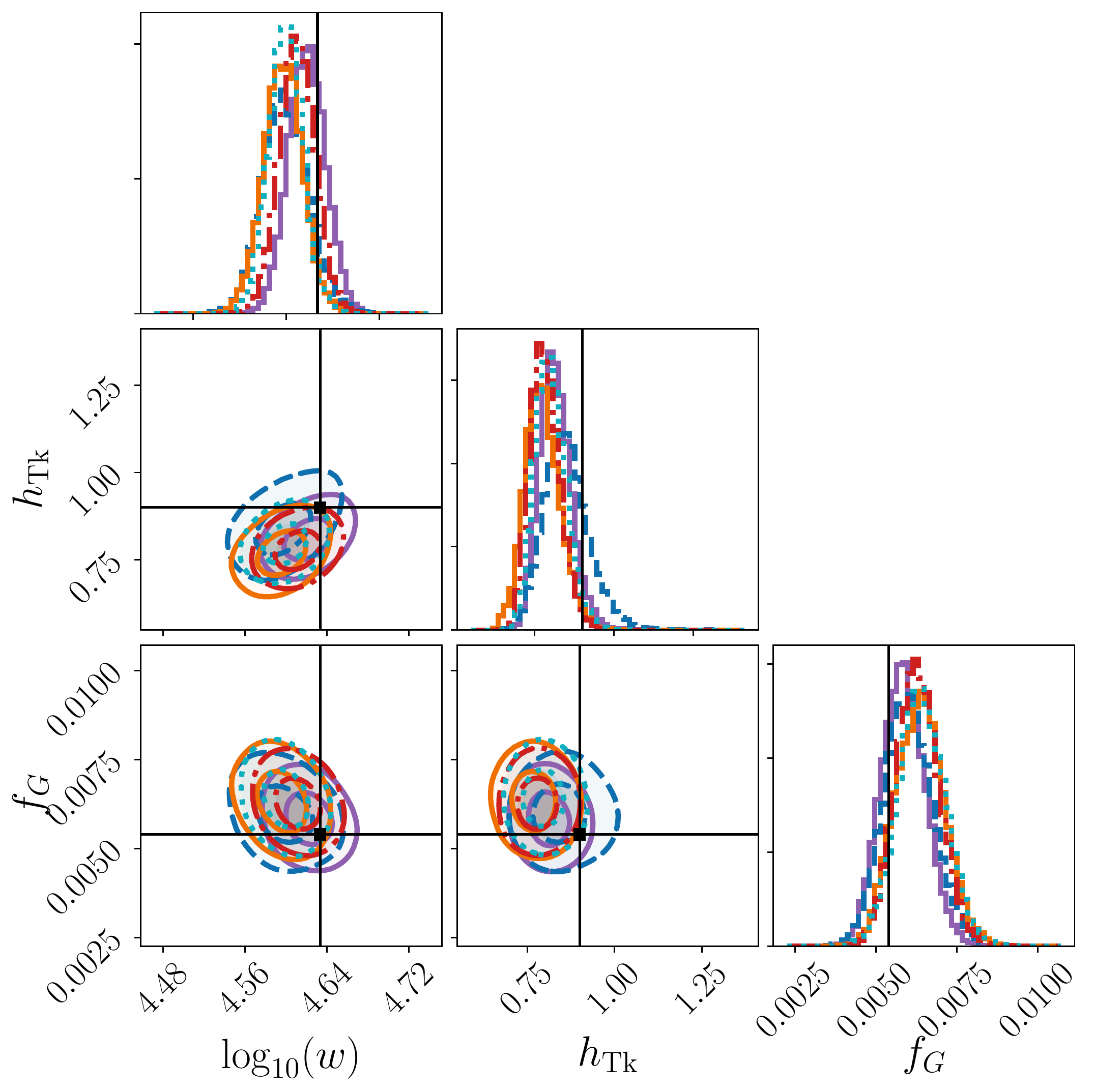}
\caption{Thick disc}
\label{fig:sys_thick_disk}
\end{subfigure}

\centering
\begin{subfigure}[b]{.73\linewidth}
\includegraphics[width=\linewidth]{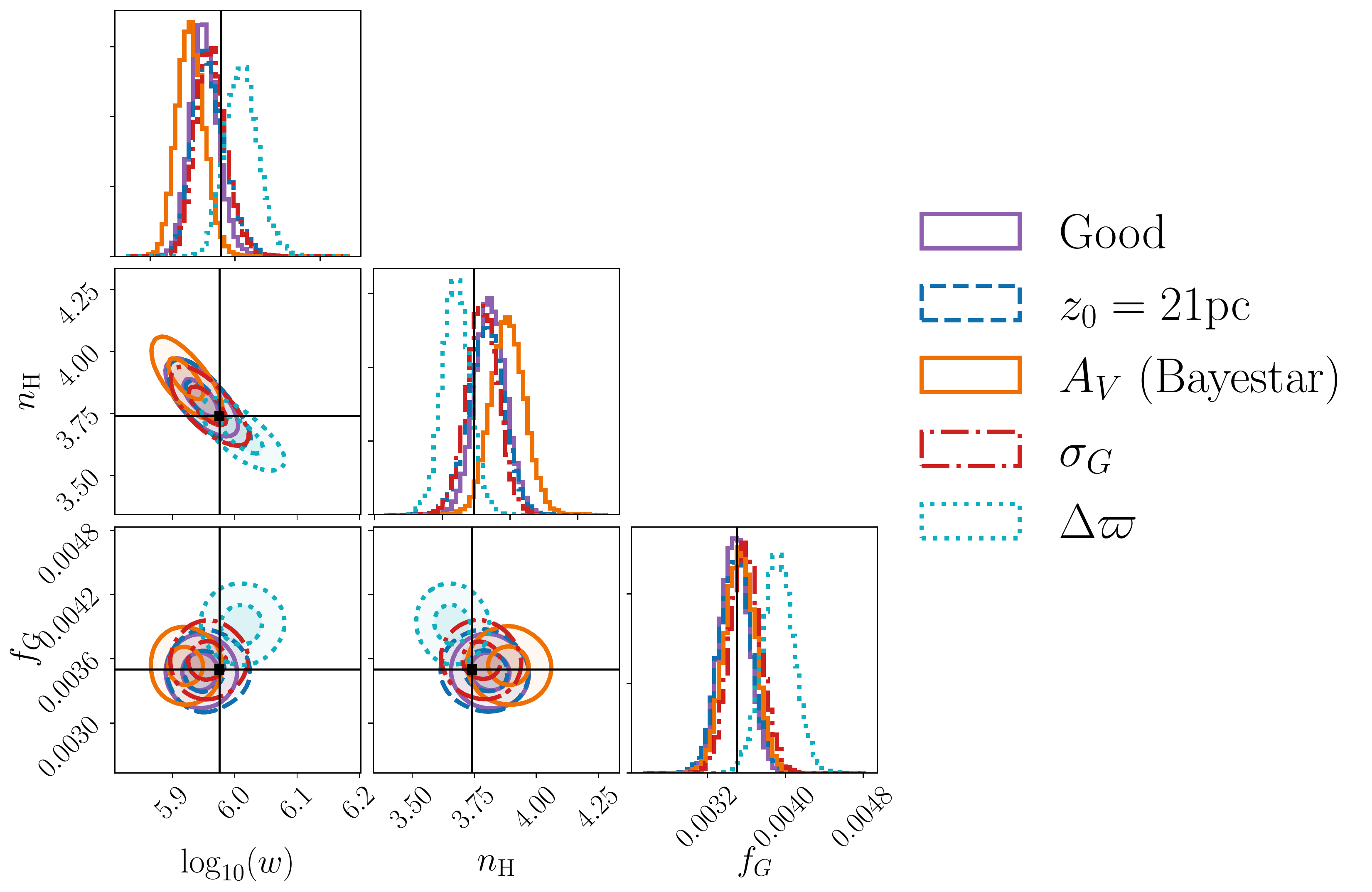}
\caption{Halo}
\label{fig:sys_halo}
\end{subfigure}

\caption{Posterior parameter fits to the mock sample from \citetalias{mwtrace1} for the thin disc parameters (a, top left), thick dick (b, top right) and halo (c, bottom) under alterations to the data which could introduce systematic errors. The purple solid ``Good'' contours in all panels show the posteriors from \citetalias{mwtrace1}, fit to the sample without any imposed systematics and black dot and lines show the input parameters used to generate the sample.  Adding a $z_\odot$ offset (blue dotted) has only a marginal impact on most parameter estimates. \gaia-like magnitude error ($\sigma_G$, red dot-dashed) leads us to underestimate the thin disc giant fraction. Extinction from Bayestar ($A_V$, orange solid) biases the model towards an overly steep halo whilst a $-10\,\mu$as parallax offset ($\Delta\varpi$, cyan dotted) has the opposite effect.}
\label{fig:sys_corners}
\end{figure*}

\renewcommand{\arraystretch}{1.5}
%\begin{center}
\begin{table*}
\begin{tabular}{c c c c c c c c} 
 \hline
 Component & Parameter & Input & Good fit & $z_0 = 21$pc & $A_V$ & $\sigma_G$ & $\Delta \varpi$ \\ [0.5ex] \hline\hline
Thin disc
 & $w$ & ${1.20}\times 10^{4}$
 & ${9.59}_{-1.35}^{+1.38}\times 10^{3}$
 & ${1.01}_{-0.17}^{+0.17}\times 10^{4}$
 & ${8.93}_{-1.55}^{+1.65}\times 10^{3}$
 & ${8.76}_{-1.27}^{+1.52}\times 10^{3}$
 & ${9.56}_{-1.37}^{+1.52}\times 10^{3}$\\
 & $h_\mathrm{Tn}$ & ${0.300}$
 & ${0.281}_{-0.015}^{+0.015}$
 & ${0.296}_{-0.018}^{+0.017}$
 & ${0.275}_{-0.018}^{+0.017}$
 & ${0.272}_{-0.015}^{+0.017}$
 & ${0.283}_{-0.015}^{+0.016}$\\
 & $f_G$ & ${4.50}\times 10^{-3}$
 & ${3.76}_{-1.30}^{+1.43}\times 10^{-3}$
 & ${3.20}_{-1.27}^{+1.37}\times 10^{-3}$
 & ${3.11}_{-1.49}^{+1.60}\times 10^{-3}$
 & ${1.67}_{-0.98}^{+1.28}\times 10^{-3}$
 & ${3.28}_{-1.36}^{+1.43}\times 10^{-3}$\\
 & $M_\mathrm{TO}$ & 3.1 & & & & & \\
 & $\alpha_3$ & -0.6 & & & & & \\
 \hline\hline
Thick disc
 & $w$ & ${4.30}\times 10^{4}$
 & ${4.19}_{-0.19}^{+0.20}\times 10^{4}$
 & ${3.97}_{-0.21}^{+0.24}\times 10^{4}$
 & ${3.95}_{-0.18}^{+0.19}\times 10^{4}$
 & ${4.09}_{-0.17}^{+0.19}\times 10^{4}$
 & ${3.98}_{-0.17}^{+0.17}\times 10^{4}$\\
 & $h_\mathrm{Tk}$ & ${0.900}$
 & ${0.812}_{-0.045}^{+0.052}$
 & ${0.846}_{-0.064}^{+0.072}$
 & ${0.772}_{-0.050}^{+0.058}$
 & ${0.780}_{-0.042}^{+0.050}$
 & ${0.798}_{-0.045}^{+0.052}$\\
 & $f_G$ & ${5.40}\times 10^{-3}$
 & ${5.83}_{-0.66}^{+0.69}\times 10^{-3}$
 & ${5.99}_{-0.72}^{+0.77}\times 10^{-3}$
 & ${6.35}_{-0.76}^{+0.74}\times 10^{-3}$
 & ${6.28}_{-0.65}^{+0.69}\times 10^{-3}$
 & ${6.40}_{-0.70}^{+0.76}\times 10^{-3}$\\
 & $M_\mathrm{TO}$ & 3.1 & & & & & \\
 & $\alpha_3$ & -0.77 & & & & & \\
 \hline\hline
Halo
 & $w$ & ${9.45}\times 10^{5}$
 & ${8.81}_{-0.45}^{+0.52}\times 10^{5}$
 & ${8.97}_{-0.51}^{+0.69}\times 10^{5}$
 & ${8.34}_{-0.43}^{+0.49}\times 10^{5}$
 & ${9.05}_{-0.51}^{+0.59}\times 10^{5}$
 & ${1.03}_{-0.07}^{+0.07}\times 10^{6}$\\
 & $n_\mathrm{H}$ & ${3.740}$
 & ${3.812}_{-0.066}^{+0.068}$
 & ${3.795}_{-0.081}^{+0.076}$
 & ${3.889}_{-0.073}^{+0.075}$
 & ${3.778}_{-0.068}^{+0.068}$
 & ${3.661}_{-0.060}^{+0.063}$\\
 & $f_G$ & ${3.50}\times 10^{-3}$
 & ${3.48}_{-0.15}^{+0.15}\times 10^{-3}$
 & ${3.49}_{-0.17}^{+0.16}\times 10^{-3}$
 & ${3.53}_{-0.16}^{+0.16}\times 10^{-3}$
 & ${3.59}_{-0.15}^{+0.16}\times 10^{-3}$
 & ${3.92}_{-0.16}^{+0.16}\times 10^{-3}$\\
 & $M_\mathrm{TO}$ & 3.1 & & & & & \\
 & $\alpha_3$ & -0.64 & & & & & \\
 \hline\hline
Shared
 & $\alpha_1$ & ${-0.110}$
 & ${-0.110}_{-0.002}^{+0.002}$
 & ${-0.110}_{-0.002}^{+0.002}$
 & ${-0.111}_{-0.002}^{+0.002}$
 & ${-0.108}_{-0.002}^{+0.002}$
 & ${-0.105}_{-0.002}^{+0.002}$\\
 & $\alpha_2$ & ${-0.250}$
 & ${-0.252}_{-0.008}^{+0.009}$
 & ${-0.252}_{-0.009}^{+0.009}$
 & ${-0.257}_{-0.009}^{+0.009}$
 & ${-0.246}_{-0.009}^{+0.009}$
 & ${-0.228}_{-0.009}^{+0.009}$\\\hline\hline 
\end{tabular}
\caption{The results of systematics due to shifting the Sun from the Galactic plane ($z_\odot=21$pc), introducing dust extinction ($A_V$), adding apparent magnitude error ($\sigma_G$) and introducing a $-10\,\mu$as zero point parallax offset ($\Delta\varpi$) are tested for the mock sample. ``Good fit'' provides the results from \citetalias{mwtrace1} where the parameters are fit to data drawn from the same model with no systematics imposed. Most tests have only a marginal impact on parameter estimates with $\lesssim 2$ standard deviation offsets from the input parameters. Magnitude error significantly affects thin disc giant fraction whilst $A_V$ and $\Delta \varpi$ have a stronger impact on the halo parameters.}
\label{tab:systematics}
\end{table*}
%\end{center}

There are various aspects of the model which may lead to systematic errors in the posterior parameter fits. In most cases, these originate from simplifications to make the optimization computationally tractable. Here, we address some of the aspects which are capable of biasing the results and test the significance of their impact on the inferred parameters.

The tests are all performed using mock catalogues. The first four systematic tests (Sections~\ref{sec:solar_offset}-\ref{sec:poff}) use the exact same sample as Section 4~of~\citetalias{mwtrace1}, but resampling from the selection function where apparent magnitudes are altered. In Fig.~\ref{fig:sys_corners} and Table~\ref{tab:systematics} we have provided the posteriors of the ``SF \& $\sigma_\varpi$'' fit from \citetalias{mwtrace1} for comparison and labeled it ``Good'' as this was fit under ideal circumstances where the data correctly represents the model. The tests in Sections~\ref{sec:mto} and \ref{sec:Rdep} use re-sampled catalogues applying the same method as Section~4~of~\citetalias{mwtrace1} including parallax uncertainties from the Astrometric Spread Function \citep{CoGIV}. There is a level of statistical error in population sampling which affects the posteriors for tests when a new catalogue is generated.

\subsection{Solar vertical offset}
\label{sec:solar_offset}

In our model, we assume that the Sun sits directly on the mid-plane of the Milky Way and as such we have a symmetric view of the Galaxy towards the north and south. In fact, the Sun is slightly vertically offset from the Galactic plane to the north by $\sim 14-21$pc \citep{Binney1997, Joshi2007, Widmark2019, Bennett2019}. As a result, our model assumes the distribution of stars in the south is closer than it actually is and in the north, too far away. This may impact the inferred scale height of the discs. 

To test the significance of this assumption, we use our mock sample and introduce a vertical shift to the effective Solar position. This is done for stars assuming the sample is entirely in the north. The vertical position is changed for all stars such that the new coordinate, $z'$ is given by
\begin{equation}
    z' = z - z_\odot,
\end{equation}
with $z_\odot=21$ pc -- towards the upper end of estimates of the Solar position offset from the Galactic plane.
This reduces both the latitude and distance of sources and therefore also reduces the apparent magnitudes
\begin{align}
    \tan b^\prime &= \frac{s\tan b - z_\odot\sec b}{s}\\
    s' &= s\left[ 1 + \frac{z_\odot}{s}\left(\frac{z_\odot}{s} - 2\sin b\right)\right]^{\frac{1}{2}}.
\end{align}
Our latitude cut is applied on the updated latitudes, $|b'|>80$. We do not use the southern population, as this requires re-sampling the mock outside the original selection bounds which would be more complicated to interpret. For the north sample, this cut simply removes some sources from the original data set. The source apparent magnitudes are then recomputed from their original absolute magnitudes and the new distances, after which the selection function is applied to the sample and finally observed parallaxes are re-sampled from the expected uncertainties. The sample size is reduced by $\sim 1.8$ per cent over the original. 

The results of the parameter fits to the new sample are given in Table~\ref{tab:systematics} and shown by the blue dashed contours in Fig.~\ref{fig:sys_corners}. The shifts of parameters from the true values are marginally significant in some cases. Specifically, the scale height of the thin disc is slightly increased which may be considered counter intuitive given that we are effectively pushing the Sun closer to sources. However, pairing this with the increased weight of the thin disc and reduced weight of the thick disc it suggests the thin disc is taking on some thick disc sources. Overall, these results suggest that the simplification to the model of setting $z_\odot=0$ is only likely to have a marginal effect on parameter estimates.

\subsection{Dust extinction}
\label{sec:dust}

Extinction due to inter-stellar dust causes stars to appear dimmer than they would otherwise be at a given distance and absolute magnitude. This is one of the motivations behind narrowing the sample to high-latitude regions. In these areas, the effects of dust extinction are small for any individual star. However, since this systematically affects all sources in the same direction, there can still be a sizable affect on the model parameter estimates.

Why don't we use the published dust maps to de-redden the sources in the \gaia samples in the first place? At first glance, this is an appealing suggestion, but there is a subtle issue here which would lead to an underestimate in uncertainties. Because the \citealt{Green2019} extinction map is evaluated using \gaia parallax information, as is our model, there would be a double counting of information. The formally correct way to handle this problem is to simultaneously fit the structure of the Milky Way and the extinction map. One immediate challenge is a strong degeneracy between extinction, distance and absolute magnitude of sources. This is a significantly more complex problem and well beyond the scope of this paper.

Nonetheless, we can gauge the impact of extinction by applying a Milky Way extinction map to the mock catalogue, re-sample the selection function from the new observed apparent magnitudes and fit the model parameters to this. 

The most detailed 3D extinction map to date for the Milky Way is that of \citealt{Green2019}. This uses apparent magnitudes from a wide range of pass-bands throughout the optical and infra-red to estimate stellar reddening, whilst \gaia parallaxes are used to provide distance information to the model. Using the \textsc{dustmaps} Python module \citep{Green2018JOSS}, we take a single sample of the extinction parameter for each source. As a proxy for the reddening vector component in the \gaia $G$ band, we use the Pan-STARRS' $g$-filter value of $3.518$ \citep[Table 1 of][]{Green2019}. The mean extinction for sources in the selected sample is $\langle \delta G \rangle \sim 0.01$.

The addition of stellar extinction from \citealt{Green2019} has a marginally significant effect on parameter estimates shown by the orange contours in the posteriors in Fig~\ref{fig:sys_corners} and more quantitatively in the $A_V$ column of Table~\ref{tab:systematics}. The scale height and normalisation of the discs and halo are pushed down and the halo is too steep. Stars further away will be more obscured by dust and less likely to be included in our sample. Our method doesn't account for this so we instead fit a marginally steeper model than is actually the case. At high latitudes, this is a small effect but if we widened our on-sky sample, this could dramatically impact the results.

\subsection{Magnitude uncertainty}

\begin{figure}
\centering
  \includegraphics[width=0.495\textwidth]{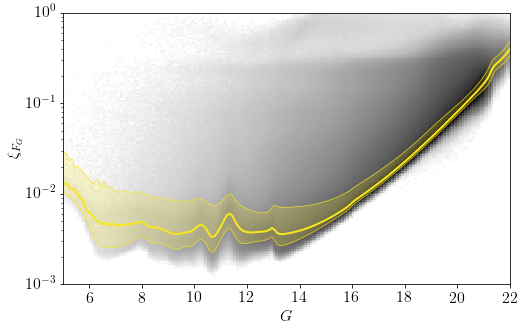}
  \caption[]{The distribution of $G$-band flux error amplitude (uncertainty per observation) of all sources in \gaia EDR3 is shown by the log-normalised grey-scale histograms. The yellow line and shaded regions provide the median and $16^\mathrm{th}-84^\mathrm{th}$ percentile range in 0.1mag bins. The median is used with the scanning law to estimate the expected apparent magnitude error for sources in \gaia as a function of position on the sky and apparent magnitude.}
   \label{fig:G_amp}
\end{figure}

The method used to fit the model assumes no apparent magnitude measurement uncertainty. Of course this is not the case and \gaia has uncertainties on all apparent magnitude measurements. To estimate the systematic effect of this uncertainty, we re-sample the mock apparent magnitudes from \gaia-like uncertainties and apply the method to the new sample. 

For any source in the \gaia catalogue, the apparent magnitude in the $G$-band is estimated up to nine times whenever it is scanned by the nine columns of CCDs in the field of view. The set of all apparent magnitude measurements, which can number in the hundreds, is used to estimate the magnitude uncertainty.
We reverse engineer this process to estimate the apparent magnitude uncertainty per observation. The amplitude of apparent magnitude measurement uncertainty is given by
\begin{equation}
    \xi_{F_G} = \sqrt{N}\frac{\sigma_{F_G}}{F_G}
    \label{eq:gamp}
\end{equation}
where $F_G$ is the measured source flux and $\xi_{F_G}$ is the flux error per observation \citep[Eq. 2][]{Belokurov2017}. We estimate this for all sources in \gaia EDR3 and take the median as a function of apparent magnitude shown in Fig.~\ref{fig:G_amp}.

To estimate apparent magnitude errors for sources in the mock sample, we then replicate \gaia's observations for those sources again. The error per observation is taken from the median in Fig.~\ref{fig:G_amp} and the number of scans of the source is given by the number of scans of that position on the sky in the EDR3 nominal scanning law\footnote{Gaia EDR3 nominal scanning law: \url{http://cdn.gea.esac.esa.int/Gaia/gedr3/auxiliary/commanded_scan_law/}}. Ideally, on average, $62/7$ observations are taken with each scan as there are nine CCD columns but in one of the seven rows, a CCD is replaced by a wave-guide sensor \citep{Prusti2016}. However, \gaia is not 100\% efficient and not all observations are successfully recorded or make it through the data processing pipeline. To account for this, we also multiply by the efficiency at a given magnitude taken from \citealt{CoGII}. Whilst this efficiency is estimated from Data Release 2 (DR2), it should give a rough approximation of the behaviour in EDR3. Eq.~\ref{eq:gamp} is now reversed to estimate the flux error
\begin{equation}
    \sigma_{F_G} = \sqrt{N(l,b,G)}\,\xi_{F_G}(G)\,F_G(G).
\end{equation}
where $N(l,b,G)$ is the product of the number of scans at the given position on the sky and $62/7$ times the observation efficiency. This is used to sample an observed flux $F_G' \sim \mathcal{N}(F_G, \sigma_{F_G})$. The mean magnitude change for all sources in our sample is $\langle |\delta G|\rangle \sim 0.003$. Finally, we apply the selection function to the newly estimated magnitudes which replicates the fact that the selection function was estimated as a function of measured apparent magnitude.

The introduction of apparent magnitude error has marginal effects on some parameter posteriors, shown by the red dot-dashed contours in Fig.~\ref{fig:sys_corners} and included in Table~\ref{tab:systematics} as $\sigma_G$, with the most significant being the thin disc dwarf fraction. This may be explained by a blurring of the sharp dwarf-giant absolute magnitude boundary which, given the low giant fraction in the thin disc, will lead to more dwarfs being estimated as giants than vice versa and reduce the dwarf fraction.

\subsection{Parallax offset}
\label{sec:poff}

As discussed in Section~\ref{sec:data}, a significant amount of work has been devoted to constraining the zero-point parallax offset of the \gaia astrometry sample. However, most tests are applied on sources at the bright end of the \gaia magnitude range. At the faint end, the correction from \citealt{Lindegren2021plx} reduces the parallax bias to $\sim$ a few micro-arcseconds as a function of apparent magnitude. However, as can be seen in the third panel of Fig. 2 in \citealt{Lindegren2021plx}, there are variations over the sky of $\sim 10\,\mu$as towards the north and south Galactic poles.

To test the impact of a residual parallax offset, we subtract $10\,\mu$as from the parallax measurements in our mock sample and rerun the fits without correcting for this. The posteriors are shown in the final column of Table~\ref{tab:systematics} and with cyan dotted contours in Fig.~\ref{fig:sys_corners}. The parallax bias has no significant impact on the thin disc and thick disc parameters however the impact on the halo is considerable. Because this is tested using the same sample as the ``Good'' fits, we are comparing with those posteriors rather than the input parameters. In the Fig.~\ref{fig:sys_halo} we can see how much the parallax offset shifts the halo parameter posteriors away from the Good results towards a shallower, more extended halo profile. This is not too surprising. A constant parallax offset is small compared to the true source parallax for nearby sources, but becomes much more significant with increasing distance. 

A similar effect was found in \citealt{Everall2019} when measuring the local velocity ellipsoid with \gaia DR2. As we stray further from the Solar neighbourhood, the impact of a negative parallax offset becomes more significant, systematically overestimating the distances to sources. In our case, this causes an overestimate of the radial extent of the halo.

% \begin{figure*}
%   \centering
%   \includegraphics[width=\textwidth]{figs/mock_sys2_sample4_1000000_028_corner.pdf}
%   \caption[]{Tests in Sections~\ref{sec:mto} and \ref{sec:Rdep} require sampling a new mock catalogue and we show the posterior distributions for these tests here compared with the `Good' fit from using the correct model for our method. Posteriors for the fit with the thin disc $M_\mathrm{TO}$, in blue dashed contours, refind the input parameters (black lines) well in all cases except $f_\mathrm{G}^\mathrm{Tn}$. The cylindrical radius-dependent sample with halo oblateness $q=0$, in orange dot-dashed contours, performs significantly worse particularly for the disc scale heights and halo power law profile.}
%   \label{fig:sys2_corner}
% \end{figure*}

\begin{figure*}

\begin{subfigure}[b]{.47\linewidth}
\includegraphics[width=\linewidth]{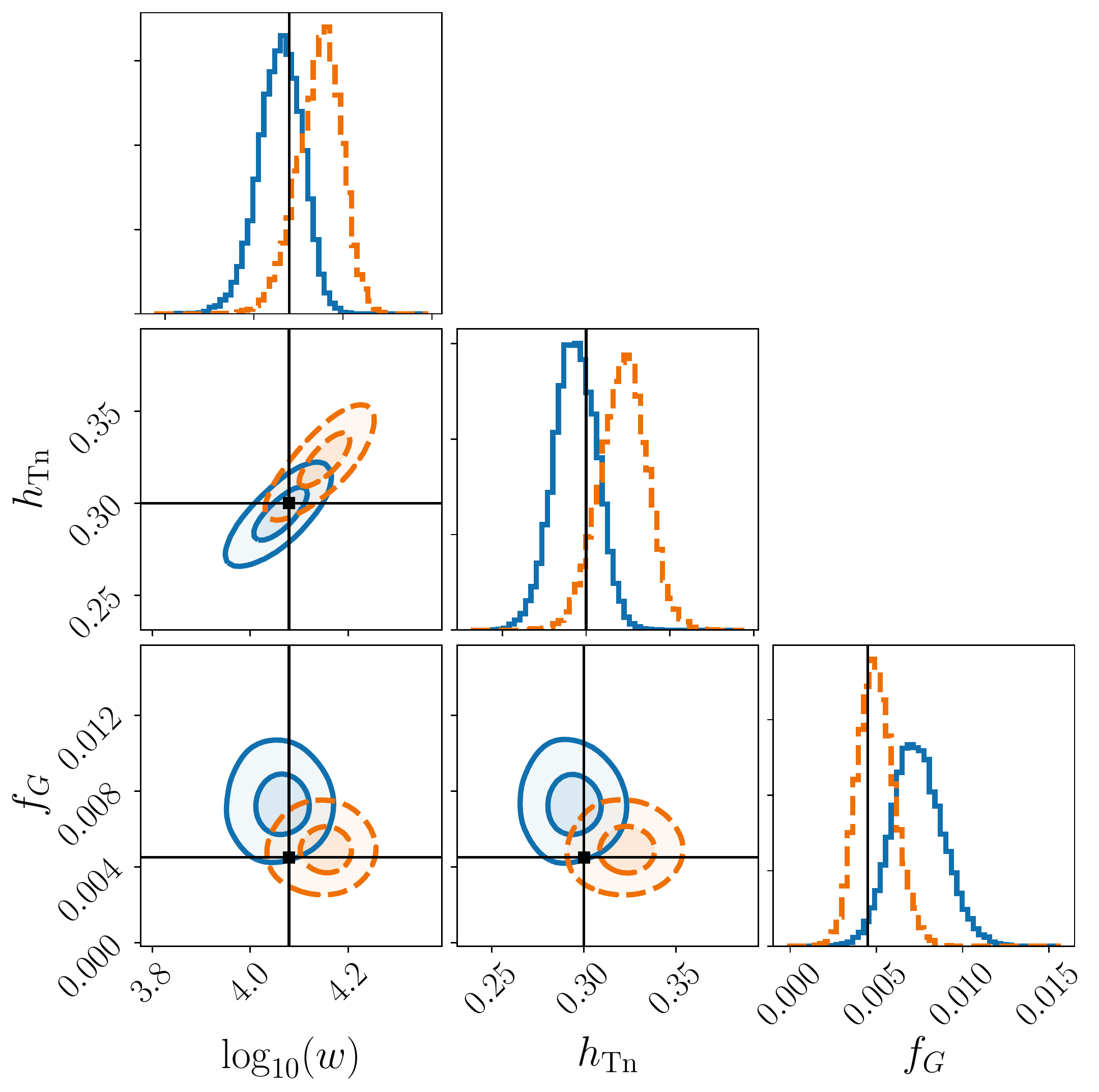}
\caption{Thin disc}
\label{fig:sys2_thin_disk}
\end{subfigure}
\begin{subfigure}[b]{.47\linewidth}
\includegraphics[width=\linewidth]{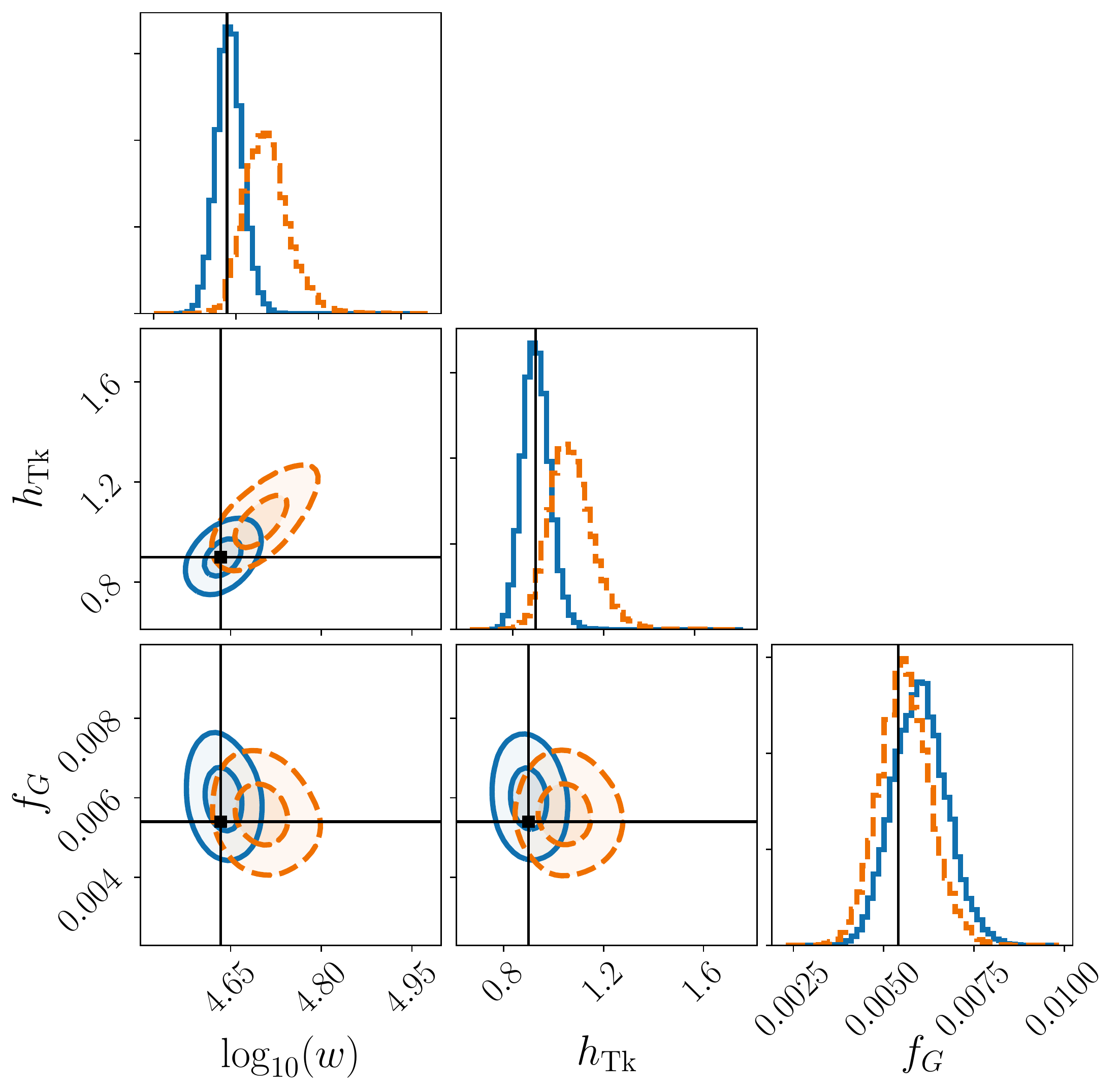}
\caption{Thick disc}
\label{fig:sys2_thick_disk}
\end{subfigure}

\begin{subfigure}[b]{.73\linewidth}
\includegraphics[width=\linewidth]{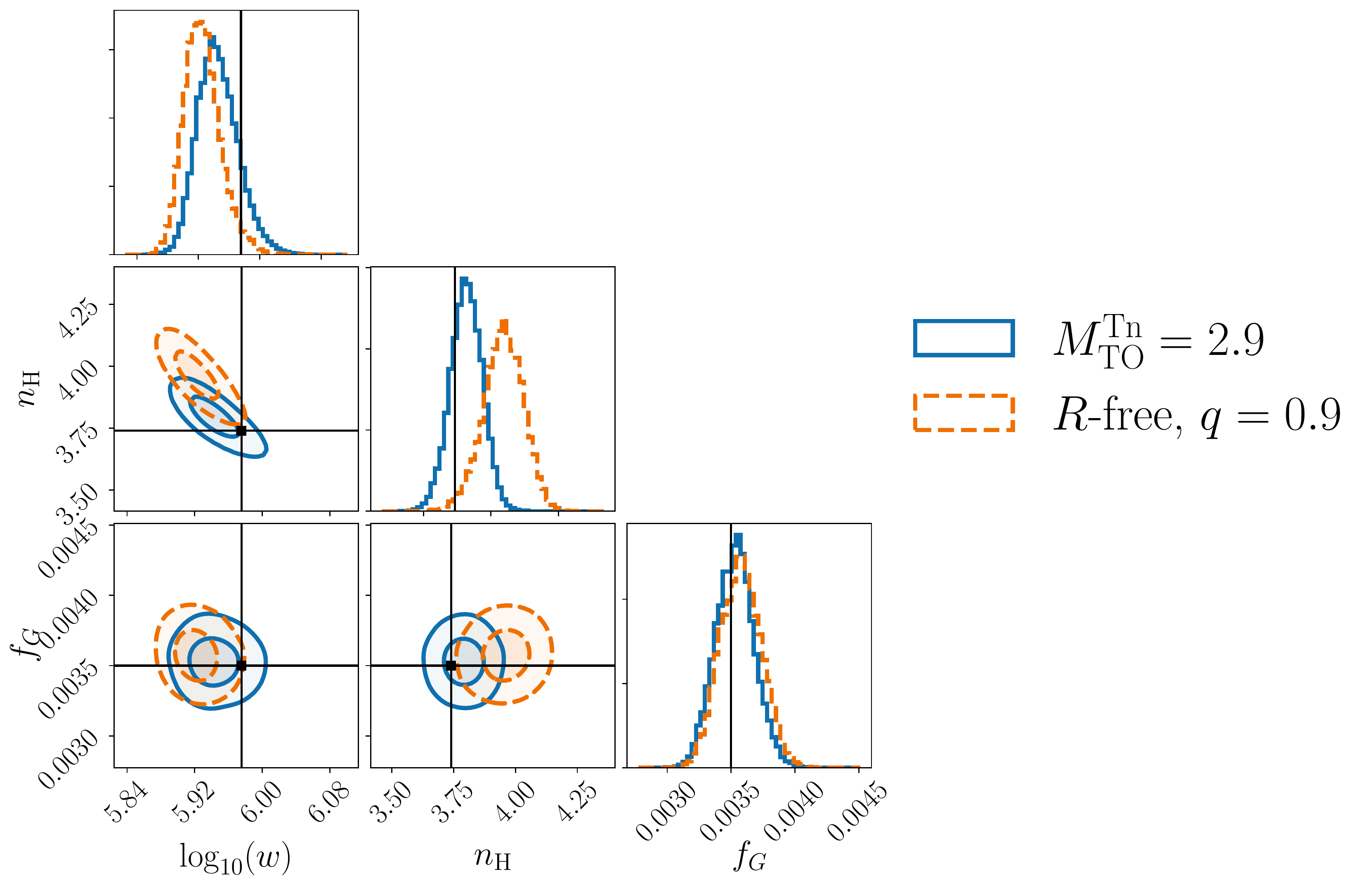}
\caption{Halo}
\label{fig:sys2_halo}
\end{subfigure}

\caption[]{Posterior distributions for parameters fit to mock samples with systematic differences to the type of model assumed in the fitting procedure. Ground truth input parameters for the mock sample are shown by the black lines and points for the thin disc (a, top left sub-figure), thick disc (b, top right sub-figure) and halo parameters (c, bottom sub-figure). The fits to the mock sample with a shifted thin disc turn-off ($M_\mathrm{TO}^\mathrm{Tn}$, blue contours) overestimate the population of thin disc giants which is unsurprising as the fits assume a giant is any source with $M_G<3.1$. Orange dashed contours are the fits to a mock sample with cylindrical radius dependence and an oblate halo ($q=0.9$, orange dashed). This significantly impacts all parameters producing extended thin and thick discs with overestimated scale heights and an overly-steep halo.
%Tests in Sections~\ref{sec:mto} and \ref{sec:Rdep} require sampling a new mock catalogue and we show the posterior distributions for these tests here compared with the `Good' fit from using the correct model for our method. Posteriors for the fit with the thin disc $M_\mathrm{TO}$, in blue dashed contours, refind the input parameters (black lines) well in all cases except $f_\mathrm{G}^\mathrm{Tn}$. The cylindrical radius-dependent sample with halo oblateness $q=0$, in orange dot-dashed contours, performs significantly worse particularly for the disc scale heights and halo power law profile.
}
\label{fig:sys2_corner}
\end{figure*}

\subsection{Turn-off magnitude}
\label{sec:mto}

When setting up the model, we fixed some parameter values. The most notable is the main-sequence turn-off which determines the absolute magnitude at which the population transitions from dwarf-dominated to giant-dominated. For all populations, we set this to $M_G=3.1$ motivated by the theoretical isochrones, however, the turn-off magnitude is a function of stellar age. We used a thin disc model with $\tau=6.9$~Gyr, but this is only approximately the mean age of the thin disc. The disc has formed over a long period of time and so is made up of sources with a wide range of ages \citep{Snaith2015, Fantin2021, Katz2021}. This leads to a main sequence turn-off which is extended over a range of magnitudes.

\citealt{RuizLara2020} demonstrated that there was a large burst of star formation in the thin disc $\sim5.9$~Gyr ago which, by inspection of isochrones, corresponds to a turn-off magnitude of $M_\mathrm{TO}=2.9$. To test this, we generate a new mock \gaia-like catalogue with the same input parameter values as described in Section~4.1~of \citetalias{mwtrace1} except that $M_\mathrm{TO}=2.9$ for the thin disc population. We then refit this incorrectly assuming a fixed $M_\mathrm{TO}=3.1$ for all components. 

The results are provided in Table~\ref{tab:systematics2} and shown by the blue solid contours in Fig.~\ref{fig:sys2_corner}. This systematic error increases the giant fraction of the thin disc, whilst not significantly affecting any other parameter. By moving $M_\mathrm{TO}$ lower and fitting with a higher value, we are classifying many dwarfs as giants in the model. Therefore, the effect on $f_\mathrm{G}$ is unsurprising but it is reassuring to see that the remaining parameters are not sensitive to small changes in the absolute magnitude distribution.

% To test how this will impact the \gaia model results, we rerun the \gaia fits with the shifted thin disc turn-off magnitude. This enables us to estimate the systematic uncertainty introduced by the fixed turn-off simplification of our model.

% The middle two columns of Table~\ref{tab:gaiaresults} show the parameter fits to the \gaia data with the thin disc turnoff set to $M_\mathrm{TO}=2.9$. By comparison with the original fits, as expected, the dwarf fraction is slightly increased for the thin disc as the dwarf-giant magnitude transition is shifted to favour dwarfs. Otherwise the only parameters which show significant shifts are the halo weight and $\alpha_2$ but notably only for the southern sample. This may be connected to the substructure of NGC 288 and SclD in the southern sample which are of course not well accounted for by the model and so may have complex effects on the parameter estimates.

\renewcommand{\arraystretch}{1.5}
\renewcommand{\tabcolsep}{5pt}
%\begin{center}
\begin{table}
\begin{tabular}{c c c c c c c c} 
 \hline
 Component & Parameter & Input & $M_\mathrm{TO}^\mathrm{Tn} = 2.9$ & $R$-free, $q=0.9$ \\ [0.5ex] \hline\hline
Thin disc
 & $w$ & ${1.20}$
 & ${1.15}_{-0.13}^{+0.13}\times 10^{4}$
 & ${1.41}_{-0.17}^{+0.15}\times 10^{4}$\\
 & $h_\mathrm{Tn}$ & ${0.300}$
 & ${0.294}_{-0.013}^{+0.013}$
 & ${0.323}_{-0.015}^{+0.014}$\\
 & $f_G$ & ${4.50}$
 & ${7.34}_{-1.39}^{+1.56}\times 10^{-3}$
 & ${4.93}_{-1.01}^{+1.12}\times 10^{-3}$\\
 \hline\hline
Thick disc
 & $w$ & ${4.30}$
 & ${4.34}_{-0.23}^{+0.25}\times 10^{4}$
 & ${5.05}_{-0.41}^{+0.49}\times 10^{4}$\\
 & $h_\mathrm{Tk}$ & ${0.900}$
 & ${0.901}_{-0.055}^{+0.061}$
 & ${1.049}_{-0.085}^{+0.099}$\\
 & $f_G$ & ${5.40}$
 & ${5.99}_{-0.73}^{+0.75}\times 10^{-3}$
 & ${5.58}_{-0.69}^{+0.71}\times 10^{-3}$\\
 \hline\hline
Halo
 & $w$ & ${9.45}$
 & ${8.78}_{-0.48}^{+0.59}\times 10^{5}$
 & ${8.39}_{-0.40}^{+0.48}\times 10^{5}$\\
 & $n_\mathrm{H}$ & ${3.740}$
 & ${3.791}_{-0.070}^{+0.070}$
 & ${3.957}_{-0.089}^{+0.086}$\\
 & $f_G$ & ${3.50}$
 & ${3.53}_{-0.15}^{+0.15}\times 10^{-3}$
 & ${3.58}_{-0.16}^{+0.16}\times 10^{-3}$\\
 \hline\hline
Shared
 & $\alpha_1$ & ${-0.110}$
 & ${-0.108}_{-0.002}^{+0.002}$
 & ${-0.111}_{-0.002}^{+0.002}$\\
 & $\alpha_2$ & ${-0.250}$
 & ${-0.249}_{-0.008}^{+0.009}$
 & ${-0.238}_{-0.008}^{+0.008}$\\\hline\hline 
\end{tabular}
\caption{We provide the median parameter estimate with $16^\mathrm{th}-84^\mathrm{th}$ percentiles for the mock sample fits with thin disc $M_\mathrm{TO}=2.9$ and the cylindrical radius dependent sample with halo oblateness $q=0.9$.}
\label{tab:systematics2}
\end{table}

\subsection{Galactocentric radius and an oblate halo}
\label{sec:Rdep}

The model used in this work has no dependence on Galactocentric cylindrical radius or azimuth. Due to the complexity of the model integration, we made the simplifying assumption that all sources have the same cylindrical radius as the Sun, $R_\odot$. For a detailed discussion of this, see Appendix~A of \citetalias{mwtrace1}.

For the thin and thick disc profiles, this means approximating the Milky Way disc as a uniform sheet. In the most extreme case, where sources are against the edge of the cone with $l=0^\circ$ or $180^\circ$, the cylindrical radius is incorrect by
\begin{equation}
    \delta R = z/\tan 80^\circ = 0.176\, z.
\end{equation}
Whilst this maximum offset is significant, it does not provide much information on how the disc profile will affect the results. For that, we examine the mean cylindrical radius offset integrating over the disc profile. Using radial scalelengths of $L_\mathrm{Tn}=2.6$~kpc and $L_\mathrm{Tk}=3.6$~kpc for the thin and thick disc respectively from \citealt{Juric2008}, we draw a sample within our $|b|>80^\circ$ cone at fixed $z$ for a thin and thick disc weighted by $\exp(-R/L)$ and estimate the mean $R$. For the thin disc, we get $\delta R \sim 0.2$~pc at $z=0.3$~kpc, whilst the thick disc produces $\delta R \sim 1.7$~pc at $z=0.9$~kpc. Therefore, the average offset of sources from their true position is small.%Therefore we have high confidence that our model is correctly measuring the vertical disc profile at the Solar radius to within the given statistical uncertainties which are typically $\gtrsim 5$ pc for the thin disc and $\gtrsim 17$ pc for the thick disc.

The halo spatial distribution is defined as a power-law profile of Galactocentric spherical radius. An incorrect cylindrical radius leads to an incorrect spherical radius. The spherical radius will be incorrect by
\begin{equation}
    \delta r = \sqrt{z^2 + R_\odot^2} - \sqrt{z^2 + y^2 + (R_\odot-x)^2},
\end{equation}
where $x,y,z$ are the standard Galactic Cartesian coordinates with $x$ positive towards the Galactic Centre. At the edges of our cone with $l=0^\circ$ and $180^\circ$ with $z=1$ kpc, this corresponds to $\delta r \sim +0.175$ and $-0.175$ kpc respectively. This increases to $\delta r \sim +1.05$ and $-1.19$ kpc for $z\sim 10$ kpc. To test the impact of this on the model fits, we sample a halo profile with $n_\mathrm{H}=3.724$ within the $b>80^\circ$ vertical cone and use the sample to estimate the mean spherical radius error. We find that the spherical radius of sources is underestimated by $\sim 0.3$ per cent on average. %This is significantly less than the statistical uncertainty of the halo profile so we expect this won't have a significant effect on the inferred halo profile.

We also assumed our halo was spherically symmetric and ignored any flattening. This assumption may impact the inferred steepness of the halo. Many works have also demonstrated that the stellar halo exhibits a tri-axial distribution \citep{Iorio2018, Iorio2019, Naidu2021}. Given the narrow vertical cone we have used, our results will primarily be sensitive to the halo oblateness parameter so we only test that in this work.% Other shape parameters of a triaxial halo may produce second order effects which are small compared to the other systematic uncertainties we are considering.}

We test the impact of placing all sources at the Solar radius and assuming a spherical halo by regenerating our mock sample with the correct cylindrical radius with an oblate halo $q=0.9$ \citep{Mateu2018}. Using the same parameters as discussed in Section 4.1 of \citetalias{mwtrace1}, we generate a mock catalogue with thin and thick disc scale lengths of $L_\mathrm{Tn}=2.6$ and $L_\mathrm{Tk}=3.6$ kpc \citep{Juric2008}. We also include the cylindrical radius dependence of the spherical radius for the halo model.

The results are shown in Table~\ref{tab:systematics2} and orange dashed contours of Fig.~\ref{fig:sys2_corner}. This has had a more significant impact on the fits with the posteriors $\sim 2-3$ sigma off the input parameter values in several cases. The disc scale heights and halo power law profile all have significant offsets from the true input parameters. It appears that the dominant effect is a level of source confusion between the components. The sample is no longer exactly representative of our assumed vertical exponential profile and power law halo but instead a marginalisation over this with radius. This leads to overestimated disc normalisation and scale heights and an overly steep halo.

\section{Statistical and Systematic uncertainties}
\label{sec:stat_and_sys}

We have produced fits to the observed data around the northern and southern Galactic poles and to mock samples to test the effects of limitations in the model. To provide results which are informative and usable, we will quantify what our results mean for the true model parameters and their statistical and systematic uncertainties.

We do this by assuming the posterior distributions for all parameters in Table~\ref{tab:transformed_results} are drawn from independent normal distributions. This enables us to parameterise all posteriors with a mean ($\mu$) and uncertainties ($\sigma$). 

The first two columns of Table~\ref{tab:transformed_results} provide the means and standard deviation uncertainties of the MCMC posteriors for the north and south fits to the \gaia data. We model the combined north/south posteriors as being drawn from a normal distribution with mean $\mu$, uncertainty $\sigma_\mathrm{sys, N/S}^2$ convolved with an additional normal distribution, $\mathcal{N}(0, \sigma^2)$ providing the standard deviation uncertainty for each sample. Therefore the likelihood of the posteriors is
\begin{align}
    \log\mathcal{L}& = \sum_{i,\mathrm{North}} \log\mathcal{N}\left(x_i\,|\,\mu, \sigma_\mathrm{North}^2 + \sigma_\mathrm{sys, N/S}^2\right) \nonumber\\
    &+  \sum_{i,\mathrm{South}}  \log\mathcal{N}\left(x_i\,|\,\mu, \sigma_\mathrm{South}^2 + \sigma_\mathrm{sys, N/S}^2\right)
\end{align}
where $x_i$ are the posterior samples provided by the MCMC chains.
We then maximise the log-likelihood with respect to $\mu$, $\sigma_\mathrm{sys,N/S}$. The results are given in the third and forth columns of Table~\ref{tab:transformed_results}. The method we are using here is similar to inflating systematics until the chi-squared reaches a `reasonable' value. However, we avoid defining an arbitrary chi-squared target by instead maximising the Gaussian log-likelihood.

Given a sample drawn from an equally weighted sum of Gaussian distributions with the same mean but different variance, the sample variance will be the mean of the individual component variances. Therefore, the statistical error for the \gaia data fits is the root-mean-square of the north and south fits
\begin{equation}
    \sigma_\mathrm{stat} = \sqrt{\frac{\sigma_\mathrm{North}^2 + \sigma_\mathrm{South}^2}{2}}.
\end{equation}
This is given as the statistical error on $\mu$ in Table~\ref{tab:transformed_results}.

We play a similar game with the results of our systematic test runs from Section~\ref{sec:systematics}. However, in this case, we know the true parameters because we provided the input parameters. The likelihood is given by
\begin{align}
    \log\mathcal{L}& = \sum_{i,\mathrm{test}} \log\mathcal{N}\left(x_i\,|\,\mu_\mathrm{true}, \sigma_\mathrm{test}^2 + \sigma_\mathrm{sys, test}^2\right) 
\end{align}
where $\sigma_\mathrm{test}$ is the statistical uncertainty of the fit given by the standard deviation of the posterior and $\mu_\mathrm{true}$ is the input parameter. We maximise this with respect to $\sigma_\mathrm{sys, test}$ to estimate the systematic uncertainty contribution from the given test. We then rescale the systematic errors by the measured Milky Way parameters, $\mu/\mu_\mathrm{true}$, to estimate the systematic error on our fits to the \gaia data. In Table~\ref{tab:transformed_results}, we provide $\sigma_\mathrm{sys, tests}$ which is the maximum systematic error for the given parameter from our tests. We also state the test(s) which dominate the systematic uncertainty contribution. Where more than one test is listed, it is because they provided a similar systematic uncertainty to within $10$ per cent.

We recommend that anyone using our results should take the root mean square sum of all quoted uncertainties to obtain the total uncertainty on each parameter.

\section{Discussion}
\label{sec:discussion}
 
Here, we interpret the results on the structural parameters of the Milky Way thin disc, thick disc and halo given in Table~\ref{tab:transformed_results},
comparing them with previous work as well as considering future developments.

\subsection{Results}

The most striking thing to notice when examining our results is the comparison between statistical and systematic uncertainties. In general, the total systematic uncertainty is more than an order of magnitude greater than statistical uncertainty. In some cases, it is over two orders of magnitude larger. This demonstrates two things. First, \gaia has ushered in an era where it is necessary to model systematic errors once considered insignificant. Rigorous systematic analysis of the kind we have performed is essential to provide accurate and reliable results. Secondly, the precision which can be achieved with \gaia data is impressive. We used a deliberately constrained sample of objects on the sky consisting of less than $0.1\%$ of the entire \gaia catalogue and yet the precision on most parameters is more than an order of magnitude better than anything in the literature.

We infer a local stellar mass density for pre-compact object stars with of ${\rho_\mathrm{local}^* = 3.66\pm0.03\,(\mathrm{stat})\pm0.52 \times10^{-2}\,\mathrm{M}_\odot/\mathrm{pc}^3\,(\mathrm{sys})}$. This is smaller than the value of
$\rho_\mathrm{local}^* \approx 4.2  \times 10^{-2}\,\mathrm{M}_\odot/\mathrm{pc}^3\,$ as derived (without errors bars) in \citealt{Flynn2006}, using the {\it Hipparcos} and {\it Tycho} surveys, together with the Catalogue of Nearby Stars).

We compute a surface density of ${\Sigma_\mathrm{local}^* = 23.17\pm0.08\,(\mathrm{stat})\pm2.43\,\mathrm{M}_\odot/\mathrm{pc}^2\,(\mathrm{sys})}$. This is significantly smaller than \citealt{Bovy2012nothick} who estimate $30\pm 1\, M_\odot/\mathrm{pc}^2$, as well as \citealt{Flynn2006} who estimate $35.5 M_\odot/\mathrm{pc}^2$. 

We expect that the most significant difference is that their works include compact objects in the stellar mass estimates. In particular, \citealt{Bovy2012nothick} use the initial mass function to infer the contribution from all sources similar to our work in Section~\ref{sec:results_mass}. However, we account for stellar mass loss and only include stars with mass low enough that they would not have evolved into a compact object or gone supernova. \citealt{Bovy2012nothick} extrapolate to higher mass stars which will have evolved to a compact object without accounting for mass loss. This means their results will significantly overestimate the total stellar mass density for evolved stars which have undergone significant mass loss. %Whilst the lack of cylindrical radius dependence does add a $\pm 2.38\,M_\odot/\mathrm{pc}^2$ systematic uncertainty to our model, this is not enough to explain the discrepancy.
We do not extrapolate our results to include compact objects as there is significant uncertainty over how much of the initial mass is kept in the final compact object remnant. 

Our relative thick-to-thin disc local density ratio sits between the values of \citealt{Mackereth2017} and \citealt{Juric2008}, although the systematic uncertainties on this due to extinction and magnitude error are quite large. 

Ample past research has been dedicated to estimating the scale heights of the thin and thick disc. There is some discrepancy between studies, with thin disc estimates in the range $h_\mathrm{Tn}\sim 120-300$pc and thick disc in the range $h_\mathrm{Tk}\sim 500-1900$pc \citep{Kuijken1989II, Bilir2006sdss, Juric2008, Ak2008, deJong2010, Mateu2018, Dobbie2020}.
We constrain the thin disc scale height as ${h_\mathrm{Tn}=260 \pm 3\, (\mathrm{stat}) \pm 26\,\mathrm{pc}\, (\mathrm{sys})}$ and thick disc ${h_\mathrm{Tk}=693 \pm 7 \,(\mathrm{stat}) \pm 121\,\mathrm{pc}\, (\mathrm{sys})}$.
Our estimates are broadly in agreement with \citealt{Juric2008}, \citealt{deJong2010} and \citealt{Mateu2018} with reasonably strong constraints on the thin disc scale height ($\pm26$ pc), but the thick disc scale height is dominated by systematic uncertainty ($\pm121$ pc) due to the cylindrical radius dependence.

The power law profile of the halo has received substantial attention with typical estimates in the range $n_\mathrm{H}\sim2.5-4.4$ \citep{Saha1985, Gould1996, Yanny2000, Newberg2006, Juric2008, Smith2009halo, deJong2010, Cohen2017, Iorio2018, Mateu2018, Hernitschek2018}. Our model sits in the middle of these estimates with $n_\mathrm{H}=3.542\pm0.023 \,(\mathrm{stat}) \pm0.259\, (\mathrm{sys})$.
Recent works have suggested the halo is better represented by a broken power law distribution \citep{Deason2011, Thomas2018, Fukushima2019}. Whilst we have focused in the inner halo by the definition of \citealt{Fukushima2019} and truncated at $s=160$ kpc, there is a wide range of distances inferred for the truncation, e.g. $25$ kpc \citep{Watkins2009} and $42$ kpc for \citep{Cohen2017}. In reality we expect we are covering both sides of the break especially considering many stars from other author's samples will likely have made it into the \gaia astrometry sample. Another issue, extensively discussed in the literature, which can impact halo fits is accreted substructure \citep{Bell2008}. Our current understanding of the halo is that it is composed of stars from from GES \citep[a major merger event $\sim 8$ Gyr ago][]{Helmi2018, Belokurov2018}, the ``Splash'' \citep[in-situ stars kicked up by the merger event,][]{Belokurov2020splash, Grand2020} and other accreted substrucuture such as Sagittarius and smaller streams. We masked problematic regions of the southern field in Section~\ref{sec:data}, however, there are likely to be more diffuse substructures which are not so easy to mask. 

Another notable feature of our results is the north-south asymmetry across several parameters. We find the northern thin disc scale height is larger than the south at $\lesssim 10$ per cent. \citealt{Dobbie2020} found a similar asymmetry, although they claimed a much larger $25$ per cent difference. We also find that the southern halo is significantly shallower with a smaller power law exponent than the north which was also seen by \citealt{Hernitschek2018}. These effects may be caused by dynamical instabilities which asymmetrically excite the disc \citep{Widrow2012, Antoja2018} and diffuse halo substructure such as Sagittarius which may contribute many more stars to the southern high latitude field \citep{Vasiliev2021sag}. 

We used our halo local mass density and profile to estimate the total halo stellar mass, using a flat uniform density for $r<1$ kpc in order to prevent the integral from diverging, obtaining $M_\mathrm{Halo}^*\sim7.2\times10^8\,\mathrm{M}_\odot$. This is quite a rough estimate of the total halo stellar mass however our results do agree reasonably well with the broad range of literature results $M_\mathrm{Halo}^* \sim 2-14\times10^8\,M_\odot$ \citep{Bell2008, Deason2011, Deason2019}.

The dwarf fractions and absolute magnitude profiles were defined specifically for this work and were mainly fit as nuisance parameters in order to get at the spatial distribution of stars so we do not discuss these in detail here.

% \subsection{Systematics}

% As we've demonstrated in Section~\ref{sec:systematics}, the impact of model simplifications on parameter results is marginal in all cases we have tested. One persistent effect we note which appears for all tests is a tendency for the thin disc to take on some weight we expect to be associated with the thick disc. This causes a systematic shift of the thin disc scale height upwards. The thick discs receives a similar effect as low $z$ thick disc stars become associated with the thin disc instead. Therefore the thick disc receives a slightly larger scale height. Finally, the power law of the halo is systematically reduced implying a flatter halo which extends further out from the Milky Way.

% \subsection{Exponential/Power-law profile}

\subsection{More general models}
\label{sec:general_models}

Our results have small statistical uncertainty compared with the dominant systematic uncertainty. This tells us that the model we have chosen to fit is over-constrained by the data. The solution to this is to significantly increase the amount of freedom in the model until our systematic and statistical uncertainties are comparable.

Some generalizations of our model are obvious: inclusion of radial dependence; provision of free parameters on the radial disc profile; allowing a free Solar vertical position $z_\odot$; introduction of halo oblateness as a free parameter. We have not provided these freedoms due to numerical complications in the parallax error integral discussed in Section~2.1 and Appendix~A of \citetalias{mwtrace1}.

We could also move away from the simplistic two-component disc model towards a continuous distribution of disc profiles
\begin{equation}
\rho(z,h_z) = \Sigma(h_z)\, \frac{\exp\left(-|z|/h_z\right)}{2h_z}
\end{equation}
where $\rho$ is the number density of stars as a function of position and scale height and $\Sigma(h_z)$ is the surface density as a function of scale height. The distinct thin and thick disc model we have applied is the simple case with
\begin{equation}
    \Sigma(h_z) = \Sigma_\mathrm{Tn}\,\delta(h_z-h_\mathrm{Tn}) + \Sigma_\mathrm{Tk}\,\delta(h_z-h_\mathrm{Tk})
\end{equation}
where $\Sigma_\mathrm{Tn}$ and $\Sigma_\mathrm{Tk}$ are the surface densities of the thin and thick discs respectively. Motivated by spectroscopic data \citet{Bovy2012nothick} proposed a model where
\begin{equation}
    \Sigma(h_z) = \mathcal{N} \exp\left(-h_z/c\right)
\end{equation}
such that the contribution of each exponential profile to the disc surface density declines exponentially with a scale length $c$. The normalisation is
\begin{equation}
    \mathcal{N} = \frac{\Sigma_\mathrm{disc}}{c\left(\exp(-h_\mathrm{min}) - \exp(-h_\mathrm{max})\right)}
\end{equation}
where $\Sigma_\mathrm{disc}$ is the total disc surface density from integrating over scale heights from $h_\mathrm{min}$ to $h_\mathrm{max}$. The total density profile as a function of $z$ generated by this scale height distribution is given by
\begin{equation}
    \nu(z) = \int_{h_\mathrm{min}}^{h_\mathrm{max}} \frac{\Sigma_\mathrm{disc}}{4 h_z c} \, \exp\left(-\frac{h_z}{c} - \frac{z}{h_z}\right) \, \mathrm{d}h_z.
\end{equation}
There is no analytic solution to this integral such that we cannot write down a simple formula for the vertical density profile. This is a significant issue for the parallax integration where our numerical integration requires a unimodal integrand as discussed extensively in \citetalias{mwtrace1}.

One further complication for this approach is that the continuous density profile is correlated with the age and metallicity of the stellar population which determine the luminosity function.

Although beyond the scope of this work, continuous disc profile models would be a worthwhile avenue to pursue, whether that entails finding forms of $\Sigma(h_z)$ which produce analytic density profiles or developing numerical methods which can fit more general models.

Alternatively, we could go for a much more data-driven approach and fit the source density at nodes with a smooth model such as a Gaussian Process to enable correlations between neighbouring points.

More ambitiously still, we could leverage the BP and RP photometry provided in \gaia EDR3 for 1.5 billion sources. Rather than using our simple magnitude model, we could directly infer the population of the HR diagram as a function of position in the Milky Way, from which the star formation histories and metallicity distributions could also be inferred. This will require selection functions for the BP and RP photometry samples which have not yet been produced.

\subsection{Extragalactic Component}

In this work, we filtered extragalactic sources from our sample using cuts on colour and excess flux. However, another option is to add an additional component to the model for sources at infinite distance.

The spatial distribution of the extragalactic sources would simply be
\begin{equation}
    \nu_{\mathrm{EG}}(l,b,s) \mathrm{d}V =  \frac{1}{2\pi(1-\sin(b_\mathrm{min}))} \,\delta(1/s)\, s^2 \, \mathrm{d}l\, \mathrm{d}\sin(b) \,\mathrm{d}s
\end{equation}
where extragalactic sources have zero parallax and are uniformly distributed across the sky for $b>b_\mathrm{min}$ or $b<-b_\mathrm{min}$ with ${b_\mathrm{min}=80^\circ}$ in this work.

However, this spatial model needs an apparent magnitude distribution for all extragalactic sources to which the selection function can be applied. This adds significantly more complexity, as the apparent magnitude distribution is dependent on the distance and luminosity distribution which are different for quasars and galaxies. For this reason, we have not chosen to model the extragalactic population in this work. However, adding this additional component would be an interesting and worthwhile route forward, measuring the population of galaxies and quasars as a function of apparent magnitude with \gaia data.

\section{Conclusions}

We used the \gaia Early Data Release 3 (EDR3) photometry and astrometry to model the vertical distribution of stars in the Milky Way at the Solar radius. Our sample includes the majority of stars with measured parallax in \gaia within $10^\circ$ of the Galactic north and south Poles. Our method formally accounts for parallax measurement uncertainty and the \gaia selection function (see the companion~\citetalias{mwtrace1} for the algorithmic details). 

We represent the vertical density of the thin and thick discs by exponentials with scale heights ${h_\mathrm{Tn}}$ and ${h_\mathrm{Tk}}$ respectively. The stellar halo density is a power-law of spherical radius, i.e., $\rho \propto r^{-n_\mathrm{H}}$. We thoroughly test possible sources of systematic uncertainty in our approach, in particular from oversimplifications of the model. This enables us to quantify the systematic uncertainty associated with all parameter estimates.

We find the scale height of the thin disc is
${h_\mathrm{Tn}=260 \pm 3\, (\mathrm{stat}) \pm 9 \pm 24\,\mathrm{pc}\, (\mathrm{sys})}$,
Here, the two levels of systematic error correspond to north-south asymmetry about the Galactic plane and simplifying model assumption (particularly the treatment of extinction and the assumption of halo spherical symmetry). The scale height of the thick disc is ${h_\mathrm{Tk}=693 \pm 7 \,(\mathrm{stat}) \pm 10 \pm 121\,\mathrm{pc}\, (\mathrm{sys})}$. Here, the larger systematic error is controlled by our 
assumption that all sources have the same cylindrical polar radius as the Sun. For the stellar halo, we are able to constrain a power law profile of $n_\mathrm{H}=3.542\pm0.023 \,(\mathrm{stat}) \pm0.160\pm0.204\, (\mathrm{sys})$. 

We infer a local stellar mass density for non-compact object stars ${\rho_\mathrm{local}^* = 3.66\pm0.03\,(\mathrm{stat})\pm0.39\pm0.34 \times10^{-2}\,\mathrm{M}_\odot/\mathrm{pc}^3\,(\mathrm{sys})}$ and surface density ${\Sigma_\mathrm{local}^* = 23.17\pm0.08\,(\mathrm{stat})\pm1.59\pm1.84\,\mathrm{M}_\odot/\mathrm{pc}^2\,(\mathrm{sys})}$. Whilst these values are lower than previous estimates \citep{Flynn2006, Bovy2012nothick}, this discrepancy may be explained by the absence of any contribution from compact object remnants to the total stellar mass. We have not included this due to the uncertain correction for stellar mass loss, itself not well accounted for in previous works.

We also find a north-south asymmetry with respect to the Galactic plane. The thin  and thick disc scale heights are larger in the north, and the halo profile is shallower in the south. However, this asymmetry is only at the $\lesssim 10$ percent level, much less than the  25 percent claimed by \citealt{Dobbie2020}.

The impressive information content of the \gaia data produces parameter estimates with significantly improved precision over previous studies, even for our sample using only a small region of the sky. However, systematics now completely dominate the error budgets meaning that we need better models to fully realise the potential of the \gaia data.

As discussed in \citetalias{mwtrace1} and Section~\ref{sec:model_recap}, the model we have applied does not represent our current understanding of the Milky Way from dynamical and chemical information. Therefore this work does not produce new insight into the physics governing the formation and evolution of the Galaxy but it does introduce a novel approach through which new insights can be found with far greater accuracy and precision than ever before.

The approach taken here demonstrates the power of information available from \gaia which has yet to be unlocked. There is a substantial prize if we can control the systematic uncertainties involved with modelling the \gaia data.
.

\section*{Acknowledgements}

AE thanks the Science and Technology Facilities Council of
the United Kingdom for financial support. DB thanks Magdalen College for his fellowship and the Rudolf Peierls Centre for Theoretical Physics for providing office space and travel funds. RG acknowledges financial support from the Spanish Ministry of Science and Innovation (MICINN) through the Spanish State Research Agency, under the Severo Ochoa Program 2020-2023 (CEX2019-000920-S).

This work has made use of data from the European Space Agency (ESA) mission \gaia (\url{https://www.cosmos.esa.int/gaia}), processed by the \gaia Data Processing and Analysis Consortium (DPAC, \url{https://www.cosmos.esa.int/web/gaia/dpac/consortium}). Funding for the DPAC has been provided by national institutions, in particular the institutions participating in the \gaia Multilateral Agreement.

AE is very grateful to Eugene Vasiliev who provided valuable assistance on numerical integration methods and who's comments on the paper content lead to many significant improvements. 

Whilst not detailed in the paper, the authors made use of the AuriGaia mock catalogues \citep{Grand2018} which helped to motivate the model and approach taken.

%%%%%%%%%%%%%%%%%%%%%%%%%%%%%%%%%%%%%%%%%%%%%%%%%%
\section*{Data Availability}

The data underlying this article are publicly available from the European Space Agency's \gaia archive (\url{https://gea.esac.esa.int/archive/}).  The selection function implementations used in this work are described in \citealt{CoGII} and \citealt{CoGV} and made publicly accessible through the Python package \textsc{selectionfunctions} (\url{https://github.com/gaiaverse/selectionfunctions}).

The code used to fit the model and produce all figures is made publicly available as a GitHub repository (\url{https://github.com/aeverall/mwtrace.git}).

%%%%%%%%%%%%%%%%%%%% REFERENCES %%%%%%%%%%%%%%%%%%

% The best way to enter references is to use BibTeX:

\bibliographystyle{mnras}
\bibliography{references}

% Alternatively you could enter them by hand, like this:
% This method is tedious and prone to error if you have lots of references
%\begin{thebibliography}{99}
%\bibitem[\protect\citeauthoryear{Author}{2012}]{Author2012}
%Author A.~N., 2013, Journal of Improbable Astronomy, 1, 1
%\bibitem[\protect\citeauthoryear{Others}{2013}]{Others2013}
%Others S., 2012, Journal of Interesting Stuff, 17, 198
%\end{thebibliography}

%%%%%%%%%%%%%%%%%%%%%%%%%%%%%%%%%%%%%%%%%%%%%%%%%%

%%%%%%%%%%%%%%%%% APPENDICES %%%%%%%%%%%%%%%%%%%%%

\appendix

\section{Surface Density Integrals}
\label{app:surfacedensity}

Here we show the maths for evaluating the surface density of each component of our model from the number density of sources.

\subsection{Discs}

The number density of sources in the each disc component is
\begin{equation}
    \nu_c = \frac{w_c\,\tan^2(b_\mathrm{min})}{2\pi\,h_c^3} \exp\left(-\frac{|z|}{h_c}\right)
\end{equation}
as given in Eq.~23 of \citetalias{mwtrace1} where subscript $c$ refers to the disc component, such that
\begin{equation}
    \int_0^{2\pi} \int_{b_\mathrm{min}}^{\pi/2} \int_0^{\infty} \nu_c\,s^2\,\cos(b)\,\mathrm{d}s\,\mathrm{d}b\,\mathrm{d}l = w_c,
\end{equation}
is the total number of sources in the cone. The surface number density is
\begin{align}
    \Sigma_c &= \int_{-\infty}^{\infty}\,\nu_c\,\mathrm{d}z \\
             &= 2\int_{0}^{\infty}\,\frac{w_c\,\tan^2(b_\mathrm{min})}{2\pi\,h_c^3} \exp\left(-\frac{z}{h_c}\right)\,\mathrm{d}z \\
             &=\frac{w_c\,\tan^2(b_\mathrm{min})}{\pi\,h_c^2}.
\end{align}
We can also write this relative to the local number density of stars 
\begin{equation}
    \nu_c(s=0)=\frac{w_c\,\tan^2(b_\mathrm{min})}{2\pi\,h_c^3}
\end{equation}
such that
\begin{equation}
    \Sigma_c = 2\,h_c\,\nu_c(s=0).
\end{equation}

\subsection{Halo}

The number density of sources in the halo is
\begin{equation}
    \nu_\mathrm{H} = \mathcal{N}_\mathrm{H} \,r^{-n_\mathrm{H}} = \mathcal{N}_\mathrm{H} \,\left(R_0^2+z^2\right)^{-n_\mathrm{H}/2}
\end{equation}
where
\begin{equation}
    \frac{2\pi\,\mathcal{N}_\mathrm{H}}{R_0^{n_\mathrm{H}-3}} \mathcal{I} = w_\mathrm{H}
\end{equation}
and $\mathcal{I}$ is the dimensionless integral
\begin{equation}
    \mathcal{I} = \int_{\sin\left(b_\mathrm{min}\right)}^{1}\int_0^{\frac{s_\mathrm{max}}{R_0}}\,x^2\,\left(1+\sin^2(b)\,x^2\right)^{-n_\mathrm{H}/2}\,\mathrm{d}x\,\mathrm{d}\sin(b).
\end{equation}
$s_\mathrm{max}$ is the upper limit of the distances used in the model as discussed in Section~\ref{sec:results}.

The surface density integral is
\begin{equation}
    \Sigma_\mathrm{H} = 2\mathcal{N}_\mathrm{H}\,\int_0^{s_\mathrm{max}} \left(R_0^2 + z^2\right)^{-n_\mathrm{H}/2} \,\mathrm{d}z.
\end{equation}
where we have made the approximation that the distance upper bound, $s_\mathrm{max}$ approximately corresponds to an upper bound on $z$ due to the narrow cone we have used. Since the number density of sources at $160$kpc is far smaller than in the solar neighbourhood, the impact of this assumption is negligible.

The surface density is then
\begin{align}
    \Sigma_\mathrm{H} &= \frac{2\mathcal{N}_\mathrm{H}}{R_0^{n-1}} \int_0^{\frac{s_\mathrm{max}}{R_0}} \left(1+x^2\right)^{-n_\mathrm{H}/2} \, \mathrm{d}x \\
    & = \frac{w_\mathrm{H}}{\pi\,\mathcal{I}\,R_0^2} \mathcal{I}_z
\end{align}
where
\begin{equation}
    \mathcal{I}_z = \int_0^{\frac{s_\mathrm{max}}{R_0}} \,\left(1+x^2\right)^{-n_\mathrm{H}/2} \mathrm{d}x.
\end{equation}
Again, we can write this in terms of the local source density
\begin{equation}
    \nu_\mathrm{H}(s=0) = \frac{w_\mathrm{H}\,R_0^{-3}}{2\pi\,\mathcal{I}}
\end{equation}
such that
\begin{equation}
    \Sigma_\mathrm{H} = 2 R_0\,\mathcal{I}_z\,\nu_\mathrm{H}(s=0)
\end{equation}
%%%%%%%%%%%%%%%%%%%%%%%%%%%%%%%%%%%%%%%%%%%%%%%%%%

% Don't change these lines
\bsp	% typesetting comment
\label{lastpage}
\end{document}